\documentclass[12pt]{article}
\usepackage{amsmath}
\usepackage{graphicx,psfrag,epsf}
\usepackage{enumerate}
\usepackage{natbib}
\usepackage{setspace}
\usepackage{algpseudocode}

\usepackage{multirow}
\usepackage{floatrow}
\newfloatcommand{capbtabbox}{table}[][1.4\FBwidth]

\usepackage{blindtext}

\usepackage[hyphens]{url}
\usepackage[hidelinks]{hyperref}
\hypersetup{breaklinks=true}
\newcommand{\blind}{0}

\usepackage{amstext,amssymb, xcolor, ulem}

\usepackage{algorithm,algpseudocode}
\newtheorem{definition}{Definition}

\newtheorem{lem}{Lemma}
\newtheorem{thm}{Theorem}

\newcommand{\norm}[1]{\left\lVert#1\right\rVert}
\newcommand{\abs}[1]{\left|#1\right|}

\begin{document}




\if0\blind
{
  \title{\bf Scalable and non-iterative graphical model estimation}
  \author{Kshitij Khare, \it{Dept. of Statistics, University of Florida}\\
    Syed Rahman, \it{Apple Inc.}\\
	Bala Rajaratnam, \it{Dept. of Statistics, UC Davis}\\
	Jiayuan Zhou, \it{Freddie Mac}}
 \date{}
  \maketitle
} \fi

\if1\blind
{
  \bigskip
  \bigskip
  \bigskip
  \begin{center}
    {\LARGE\bf Scalable and non-iterative graphical model estimation}
\end{center}
  \medskip
} \fi

\bigskip
\begin{abstract}
Graphical models have found widespread applications in many areas of modern statistics and machine learning. Iterative Proportional Fitting (IPF) and its variants have become the default method for undirected graphical model estimation, and are thus ubiquitous in the field. As the IPF is an iterative approach, it is not always readily scalable to modern high-dimensional data regimes. In this paper we propose a novel and fast non-iterative method for positive definite graphical model estimation in high dimensions, one that directly addresses the shortcomings of IPF and its variants. {\color{black} In addition, the proposed method has a number of other attractive properties.} {\color{black} First, } we show formally that as the dimension $p$ grows, the proportion of graphs for which the proposed method will outperform the state-of-the-art in terms of computational complexity and performance tends to 1, affirming its efficacy in modern settings. {\color{black} Second,} the proposed approach can be readily combined with scalable non-iterative thresholding-based methods for high-dimensional sparsity selection. {\color{black} Third,} the proposed method has high-dimensional statistical guarantees. Moreover, our numerical experiments also show that the proposed method achieves scalability without compromising on statistical precision. {\color{black} Fourth, unlike the IPF, which depends on the Gaussian likelihood, the proposed method is much more robust.}

\end{abstract}

\noindent%

\setstretch{1}
\section{Introduction}
\label{sec:intro}

Sparse covariance estimation and graphical models have become a staple of modern statistical inference and machine learning. They have also found widespread use in a spectrum of application areas. In this paper we specifically focus on the large class of undirected (or concentration) graph models, where some entries of the inverse covariance matrix $\Omega$ are restricted to be zero. These models have been of significant interest in recent years, especially as high-dimensional datasets with {\color{black} tens of thousands} of variables {\color{black}or more} have become increasingly common. The cases when the underlying distribution is Gaussian has also been of special interest. Under this assumption, zeroes in the concentration matrix imply conditional independencies.

Suppose ${\bf Y}_1, {\bf Y}_2, \cdots, {\bf Y}_n$ are i.i.d. observations from a distribution on $\mathbb{R}^p$. Let $\Sigma$ denote the covariance matrix, and let $\Omega = \Sigma^{-1}$ denote the concentration matrix or the inverse covariance matrix of this distribution. There are often two tasks involved in the statistical analysis of undirected graphical models. The first is {\it model or graph selection}, or identifying the pattern of structural zeros in $\Omega$. The second is {\it positive definite estimation}, i.e., finding a positive definite (p.d.) estimate of $\Sigma$ under the discovered sparsity pattern. Such positive definite estimates are often required for downstream applications where the covariance estimate is a critical ingredient in multivariate analysis techniques. Examples of such applications are widespread in the biomedical, environmental and social sciences.

A number of useful methods have been proposed for positive definite inverse covariance matrix estimation consistent with a given sparsity pattern. For Gaussian undirected graphical models, the MLE cannot be obtained in closed form unless the pattern of zero restrictions corresponds to a decomposable or chordal graph (see \cite{lrtzngphmd}). {A novel iterative algorithm called iterative proportional fitting (IPF) was proposed by \cite{speedkvrgr} to obtain the MLE in this case. Over the last few decades, the IPF estimation method has become synonymous with undirected graphical model estimation for a given sparsity pattern. Several faster modifications and adaptations of IPF have been proposed in recent years, highlighting the contemporaneous nature of this line of research, see for example \cite{Hara:Takemura:2010,XGH:2011,XGT:2012,XGT:2015,Hogsgard:Lauritzen:2023}. These iterative algorithms rely on sequential updates to relevant sub-matrices of $\Sigma$ or $\Omega$ based on the given sparsity pattern.

A different approach which obtains the MLE by modifying the Glasso algorithm to take the known sparsity pattern into account has been developed in \cite[Algorithm 17.1]{ESL}. The likelihood is maximized one row at a time using block coordinate-wise ascent. We refer to this algorithm as the {G-IPF} algorithm.} There is also substantial literature on Bayesian estimation for Gaussian undirected graphical models, see \cite{Dawid:Lauritzen:1993, Roverato:2002, Letac:Massam:2007, rmcfcestgm, MME:2011, Lenkoski:2013} and the references therein. Numerical approaches for estimation in undirected graphical models such as the IPF and its variants, and the G-IPF, all tend to employ computationally intensive iterative algorithms which have to be repeated until convergence. The iterative nature of these algorithms thus does not always render them immediately scalable to the modern setting where scalability is a real necessity. Such modern regimes pose a significant challenge in the sense that they require scalability without compromising on statistical precision.

As mentioned above, the iterative nature of current methods is an immediate impediment to scalability, in the sense that numerical computations of a high order have to be repeated, thus resulting in a prohibitive computational burden. Hence, it is critical to construct a possibly {\it non-iterative method} which is able to provide a {\it positive definite estimate of $\Omega$ which is consistent with a given general sparsity pattern}. A new estimation framework, one which is specifically designed for and addresses the high dimensional nature of the problem, is required. We address this crucial gap in the literature by introducing a non-iterative method, called {CCA}, for positive definite estimation in undirected graphical models with an {\bf arbitrary} sparsity pattern in the concentration matrix. The method is comprised of two steps and can be summarized as follows: In the first step, estimates are obtained for a pattern of zeros corresponding to an appropriate larger chordal graph. As the approximating graph is chordal, estimates can be immediately computed in closed form, circumventing repeated iterations. In the second step, the Cholesky decompositions of such estimates are then suitably modified to obtain estimates with the original pattern of zeroes in the concentration matrix. These steps together yield an approach which obviates the need to employ computationally expensive iterative algorithms when dealing with non-chordal concentration graphical models. Specifically, the chordal graph serves as an important conduit to obtaining the desired concentration graph estimates. {\color{black} Notably, methods such as the IPF and its modern variants require repetition of its iterations until convergence, while no such repetition is necessary for the {CCA} algorithm. Thus, the proposed {CCA} algorithm provides significant computational savings compared to IPF or {G-IPF}. Moreover, in several settings the computational complexity of the {CCA} algorithm is smaller than even a single iteration of the {G-IPF} algorithm (see Section \ref{sec:complexity} for examples), bearing in mind that the G-IPF iterations have to be repeated thereafter till convergence.} We demonstrate both theoretically and numerically that substantial computational improvements can be obtained using the proposed approach, while maintaining similar levels of statistical precision as state-of the art approaches.

{\color{black}In addition to the introduction of a novel non-iterative estimation framework, the proposed CCA approach has multiple other benefits - from both a theoretical and methodological perspective. Recall that the IPF and its variants fundamentally depend on the functional form of the Gaussian likelihood. On the contrary, the proposed CCA approach can instead also be formulated in a non-parametric way yielding a more robust estimator, thus making it more immune to the effect of outliers. CCA also enjoys high dimensional statistical guarantees. The proposed CCA approach has yet another significant benefit} in the sense it can be readily coupled with any scalable thresholding method for obtaining a sparsity pattern/ pattern of zeros in the inverse covariance matrix. More specifically, non-iterative thresholding based methods for model selection, such as {\color{black} the HUB screening method in \cite{Hero:Rajaratnam:2012}, and the recently proposed {FST} method in \cite{zhang2021quadratic}}, are equipped with high-dimensional statistical guarantees and are generally much faster and scalable to significantly higher dimensional settings than the penalized likelihood methods which use iterative optimization algorithms. To clarify, the term non-iterative sparsity selection method refers to a method which requires a fixed and finite amount of time, and does not need repetitions until convergence. However, such thresholding based methods, though very useful, are not guaranteed to produce positive definite estimates of $\Omega$, {\color{black} which in turn could hinder their use in} downstream applications which require a positive definite estimate of the covariance matrix {\color{black}(see for example \cite{Ledoit:Wolf:2003, GRE:2015})}. Coupling the proposed CCA approach with scalable thresholding methods however guarantees computational efficiency and scalability of the whole procedure as well as positive definiteness of the final inverse covariance matrix estimate. In principle, the proposed CCA approach can also be combined with $\ell_1$-penalized methods: Recall that $\ell_1$-penalized methods such as CONCORD (\cite{KOR:2015}) yield sparse inverse covariance matrix estimates but are not guaranteed to yield positive definite estimates. Here, one can employ the proposed CCA approach to produce a p.d. estimator which is consistent with a sparsity pattern induced by any $\ell_1$-penalized method. 

The paper is organized as follows. Section \ref{sec:preliminaries} provides required preliminaries. The proposed positive definite estimation method is introduced in Section \ref{sec:congraphmd}. A thorough evaluation of its computational complexity is provided in Section \ref{sec:complexity}. In Section \ref{sec:asymptotics} we establish asymptotic consistency of the proposed estimates in the high-dimensional setting where the number of variables increases with sample size. Detailed experimental evaluation, validation and applications to real data is undertaken in Section \ref{sec:experiments}.

\section{Preliminaries} \label{sec:preliminaries}

\noindent
In this section we provide the required background material from graph theory and matrix algebra. 

\subsection{Graphs and vertex orderings}

\noindent
A graph $G = (V,E)$ consists of a finite set of vertices $V$ (not necessarily ordered) with $|V| =: p$, and a 
set of edges $E \subseteq V \times V$. We consider undirected graphs with no loops. Hence, 
$(v,v) \notin E$ for every $v \in V$, and $(u,v) \in E \Rightarrow (v,u) \in E$. Next, we define the 
notions of vertex ordering and ordered graphs. These are necessary in subsequent analysis. 
\begin{definition} (Vertex ordering)
	A {\it vertex ordering} $\sigma$ of the vertex set $V$ is a bijection from $V$ to $V_p := 
	\{1,2, \cdots, p\}$. 
\end{definition}

\begin{definition} (Ordered graph)
	Consider the undirected graph $G = (V,E)$ and an ordering $\sigma$ of $V$. The {\it ordered graph} $G_\sigma$ has vertex set 
	$V_p$ and edge set $E_\sigma = \{(\sigma(u), \sigma(v)): (u,v) \in E\}$. 
\end{definition}

\subsection{Cholesky decomposition}

\noindent
Given a positive definite matrix $\Omega \in \mathbb{P}_p^+$, there exists a unique lower triangular matrix $L$ with positive 
diagonal entries such that $\Omega = L L^T$, 
where $L$ is the ``Cholesky factor" of $\Omega$. The 
Cholesky decomposition will be a useful ingredient in the estimators constructed in this paper. 

\subsection{The spaces $\mathbb{P}_{G_\sigma}$ and $\mathbb{L}_{G_\sigma}$}

\noindent
Let $G = (V,E)$ be a graph on $p$ vertices and let $\sigma$ be an ordering of $V$. Let 
$\mathbb{P}^+$ denote the space of $p \times p$ positive definite matrices, and 
$\mathbb{L}^+$ denote the space of $p \times p$ lower triangular matrices with positive 
diagonal entries. We consider two matrix spaces that will play an important role in our 
analysis. The first matrix space 
$$
\mathbb{P}_{G_\sigma} = \left\{ \Omega \in \mathbb{P}^+: \Omega_{ij} = 0 \mbox{ if } i \neq j \mbox{ and } (i,j) \notin 
E_\sigma \right\}, 
$$

\noindent
is the collection of all $p \times p$ positive definite matrices where the $(i,j)^{th}$ entry 
is zero if $i$ and $j$ do not share an edge in the ordered graph $G_\sigma$. The other matrix space 
$$
\mathbb{L}_{G_\sigma} = \left\{ L \in \mathbb{L}^+: LL^T \in \mathbb{P}_{G_\sigma} \right\}, 
$$

\noindent
is the collection of all $p \times p$ lower triangular matrices with positive diagonal 
entries, which can be obtained as a Cholesky factor of a matrix in $\mathbb{P}_{G_\sigma}$. 

\subsection{Decomposable graphs}

\noindent
An undirected graph $G$ is said to be {\it decomposable} if it is connected and does not contain a cycle of 
length greater than or equal to four as an induced subgraph. The reader 
is referred to \cite{lrtzngphmd} for all the common notions of graphical models (and 
in particular decomposable graphs) that we will use here. One such important notion is that of 
a perfect vertex elimination scheme. 
\begin{definition}
	An ordering $\sigma$ of $V$ is defined to be a {\it perfect vertex elimination scheme} for a decomposable graph $G$ if 
	for every $u,v,w \in V$ such that $\sigma(u) > \sigma(v) > \sigma(w)$ and $(\sigma(u), \sigma(w)) \in E_\sigma, (\sigma(v), 
	\sigma(w)) \in E_\sigma$, then $(\sigma(u), \sigma(v)) \in E_\sigma$. 
\end{definition}

%
%
%
%

\subsection{Cholesky factors and fill-in entries} \label{sec:cholfillin}

\noindent
Let $G$ be a decomposable graph, and $\sigma$ an ordering of $V$. \cite{plsnpwrsmh} 
prove that $\sigma$ is a perfect vertex elimination scheme for $G$ if and only if 
$$
\mathbb{L}_{G_\sigma} = \left\{ L \in \mathbb{L}^+: L_{ij} = 0 \mbox{ if } i > j, (i,j) \notin E_\sigma 
\right\}, 
$$

\noindent
i.e, the structural zeroes in every matrix in $\mathbb{P}_{G_\sigma}$ are also reflected in its 
Cholesky factor. 

\indent
However, if the graph $G$ is non-decomposable, or if $G$ is decomposable but $\sigma$ is not 
a perfect vertex elimination scheme, then for any matrix in $\mathbb{P}_{G_\sigma}$, the zeroes in the 
Cholesky factor are in general a subset of the structural zeroes in $\mathbb{P}_{G_\sigma}$. The 
extra non-zero entries in the Cholesky factor are refered to as the {\it fill-in} entries 
corresponding to $\mathbb{P}_{G_\sigma}$. For example, consider the non-decomposable $4$-cycle graph $G = (V,E)$ below where $\sigma$ is such that $E_\sigma = \{(1,2),(2,3),(3,4),(4,1)\}$ with $\Omega$ and $L$ specified as below. 
\begin{equation}
\vcenter{\hbox{\includegraphics[width=1.5in,height=1.5in]{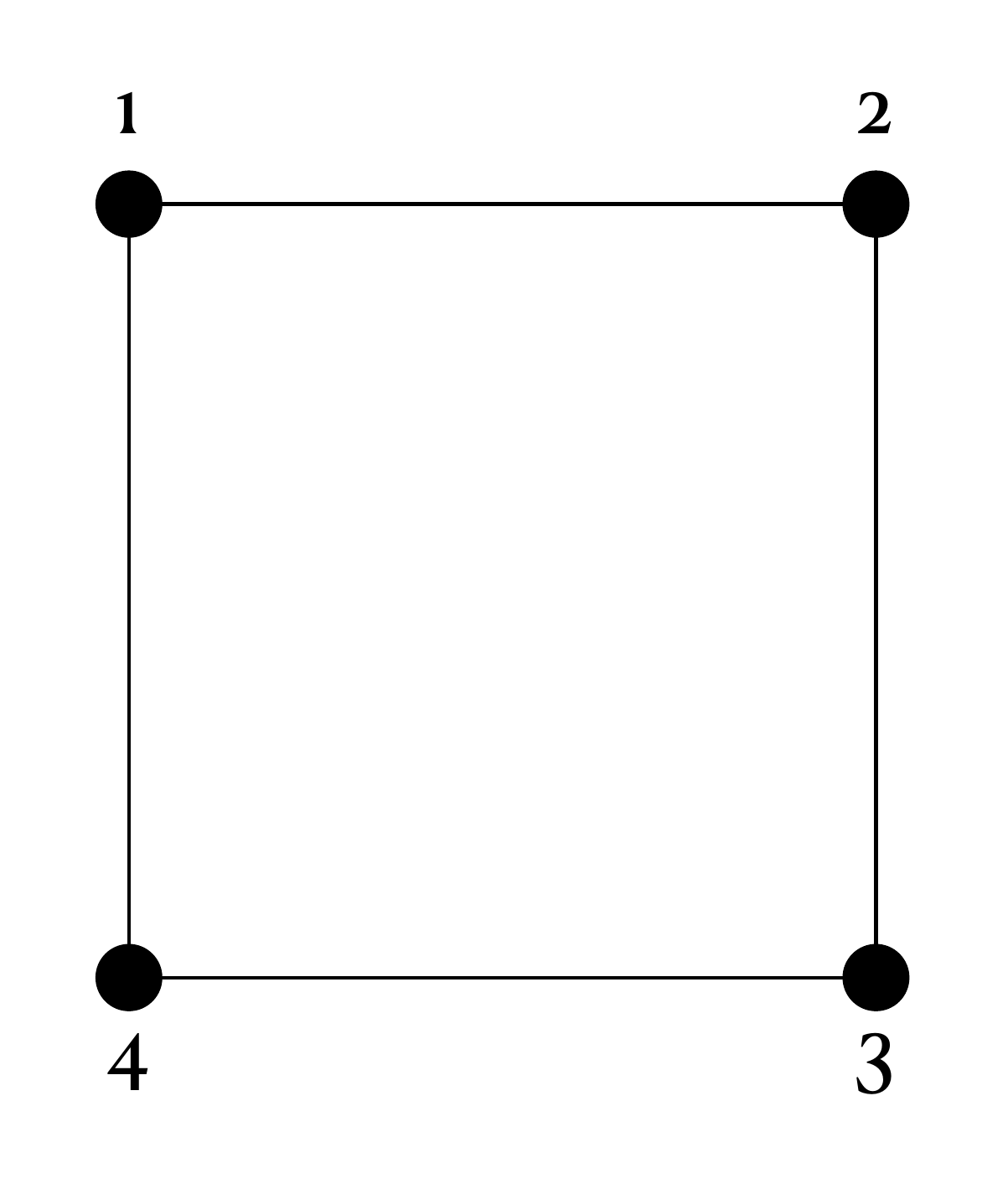}}}, \; 
\Omega = \left( \begin{matrix} 
3 & 1 & 0 & 1 \cr 
1 & 3 & 1 & 0 \cr 
0 & 1 & 3 & 2 \cr 
1 & 0 & 2 & 3 \cr 
\end{matrix} \right) \hspace{0.1in} \mbox{ and } \hspace{0.1in}
L = \left( \begin{matrix} 
1.732 & 0 & 0 & 0 \cr 
0.577 & 1.633 & 0 & 0 \cr 
0 & 0.612 & 1.620 & 0 \cr 
0.577 & -0.204 & 1.312 & 0.951 \cr 
\end{matrix} \right). 
\end{equation}

\noindent
Note that $\Omega \in \mathbb{P}_{G_\sigma}$, and it can be shown that 
$L$ above is its Cholesky factor. Clearly, the zero entries in the lower triangle of $L$ are a strict subset of the zero entries in $\Omega$. The fill-in entry in this case is the $(4,2)^{th}$ entry. Next, we define the notion of a filled graph. 
\begin{definition}
	Given an ordered graph $G_\sigma$, the {\it filled graph} for $G_\sigma$, denoted by $G_\sigma^D = (V_p, E_\sigma^D)$, is 
	defined as follows: 
	\begin{enumerate}
		\item $G^D_\sigma$ is an ordered decomposable graph, i.e., $G_D$ is a decomposable graph and 
		$\sigma$ is a perfect vertex elimination scheme for $G_D$. 
  \item For every $\Omega \in \mathbb{P}_{G_\sigma}$, the corresponding Cholesky factor $L$ satisfies 
		$L_{ij} = 0$ if $i > j, (i,j) \notin E_\sigma^D$. 
		\item The edge set $E_\sigma^D$ of $G_\sigma^D$ is a subset of the edge set of every graph 
		satisfying properties $1$ and $2$. 
	\end{enumerate}
\end{definition}

\noindent
Essentially, the filled graph for $G_\sigma$ is the ``minimal" (ordered) decomposable graph whose 
edge set contains $E_\sigma$ along with all the edges corresponding to the fill-in entries for 
matrices in $L_{G_\sigma}$, i.e., a ``decomposable cover" for $G_\sigma$. In the $4$-cycle example considered above, the 
filled graph $G_\sigma^D$ has edge set given by 
$$
E_\sigma^D = \{(1,2),(2,3),(3,4),(4,1),(4,2)\}. 
$$

\noindent
For given $G$ and $\sigma$, a simple procedure to obtain $G_\sigma^D$ called the elimination tree method (see \cite{davisdrmsl}) is given below. 
\begin{enumerate}
	\item Start with $E_\sigma^D = E_\sigma$. Set $k = 1$. 
	\item Let $\mathcal{N}_k^{>} = \{j: j > k, (j,k) \in E_\sigma^D\}$. Add necessary edges to $E_\sigma^D$ so 
	that $\mathcal{N}_k^>$ is a clique. 
	\item Set $k = k+1$. If $k < p-1$, go to Step $1$. 
\end{enumerate}

\subsection{Choice of the ordering $\sigma$} \label{sec:choiceofordering}

\noindent
The proposed CCA method presented in the next section will require Cholesky decompositions of appropriate positive definite 
matrices. Since the Cholesky decomposition inherently uses an ordering of the variables, the choice of this 
ordering is an important issue which needs to be addressed in a principled way. In several applications, a 
natural domain-specific ordering of the variables based on time, location etc. is available (see 
examples in \cite{HLPL:2006, rmcfcestgm, Shojaie:Michailidis:2010, Yu:Bien:2016, KORR:2019}). However, in other 
applications a natural ordering is not always available. In such settings, we can choose the ordering of the variables 
based on computational considerations as follows. Note that the decomposable cover $G_\sigma^D$ is a function of the ordering $\sigma$. As we shall see in Section \ref{sec:congraphmd}, it is important to choose $\sigma$ which gives as fewer fill-in entries as possible, i.e., the number of extra edges in $E_\sigma^D$ as compared to $E_\sigma$. 
This problem is well-studied in the computer science literature. In fact, it is an NP-hard problem 
which makes it challenging to solve in practice (see \cite{davisdrmsl} and the references therein). However, 
several heuristic methods have been developed to solve this problem, including the Reverse Cuthill-Mckee (RCM) method, which is implemented in the {\it chol} function in R. We refer to the ordering produced by this method as the 
{\it RCM fill reducing ordering}. In subsequent analysis, we choose $\sigma$ as the RCM ordering in case a natural ordering of the variables is not available.

\section{A non-iterative Gaussian graphical model estimator} \label{sec:congraphmd}

\subsection{Methodology: The constrained Cholesky approach}

\noindent
Suppose ${\bf Y}_1, {\bf Y}_2, \cdots, {\bf Y}_n$ are i.i.d. observations from a distribution on $\mathbb{R}^p$ with covariance matrix $\Sigma$ and concentration matrix $\Omega = \Sigma^{-1}$. The sparsity pattern in $\Omega$ is 
assumed to be given, as encoded by the graph $G$. When the distribution is multivariate Gaussian, the zeros in $\Omega$ signify conditional independence between the relevant variables, and $G$ can be interpreted as a {\it conditional independence graph}. Even when the underlying distribution is not Gaussian, sparsity in $\Omega$ signifies the lack of relevant partial correlations, and $G$ can be interpreted as a {\it partial correlation graph}. {\color{black}In several applications, $G$ is not available based on domain-specific knowledge, in which case we use an estimate of $G$ provided by a thresholding based approach such as {\it FST} (\cite{zhang2021quadratic}). As specified earlier, similar to the IPF our goal is to find a p.d. estimate of $\Omega$ consistent with a given sparsity pattern, but one that is scalable to modern data regimes. For a given $\sigma$ (or one obtained by RCM fill reducing approach), the task considered in this sub-section is to find a positive-definite estimate of $\Omega$ lying in 
$\mathbb{P}_{G_\sigma}$, without recourse to iterative approaches. 

Before we formally describe the specific aspects of the CCA methodology, we first provide an intuitive description of the two steps which encapsulate the proposed CCA method. Recall that graphical model estimation for a general non-decomposable graph faces three significant challenges (especially in high-dimensional settings): (1) lack of a closed form solution warranting an iterative procedure (such as the IPF), (2) the graphical model constraint where estimates need to lie in the correct parameter space (i.e., the prescribed zero pattern), and (3) the more nuanced and ``delicate" positivity requirement (note that the sparsity and positivity constraints constitute two distinct hurdles).

CCA specifically recognizes the above three challenges and provides a solution which overcomes all three of them in a computationally tractable manner. The order in which this is undertaken is crucial. First, CCA circumvents the need for an iterative procedure by transferring the problem from the given graph G to a decomposable cover of G (as denoted by $G^D$). Doing so, ensures that a closed form solution is obtained for a problem which is ``close" to the original estimation problem. Transferring the problem to a decomposable cover, however, introduces additional edges in the graph $G^D$ which are not present in the original graph $G$, i.e., some of the required zeros in the precision matrix estimator are lost (the additional parameters in $\hat{\Omega}^D$ are referred to as ``fill-in entries", the reason for which will become clear very soon). More specifically, the issue with the resulting decomposable graph estimator is that it does NOT have the correct sparsity pattern - though it is positive definite. As positivity is a more ``delicate" property, it can be easily lost if one zeros out the resulting decomposable estimator to achieve the desired sparsity pattern. In anticipation of the risk of losing positivity while aiming for the correct sparsity pattern, the resulting decomposable graph estimator is first expressed in terms of its Cholesky decomposition (which is available in closed form thus avoiding any iterative procedure). There are two reasons for this. The first reason is that if the correct vertex ordering is used, the zeros in $\hat{\Omega^D}$ are preserved. The second reason is that the Cholesky entries do not have to obey any positivity constraints and are therefore much easier to work with. It is thus an ideal setting to work in to try and obtain the desired sparsity pattern since deforming a Cholesky does not affect positivity (provided the diagonals are positive). In summary, obtaining the decomposable cover estimator and its corresponding Cholesky transform constitute Step 1 of the CCA approach.

Armed with the closed form Cholesky estimator for the decomposable graph, the remaining task is to obtain the correct sparsity pattern for the final estimate of $\Omega$, i.e., to ensure that the final estimate lies in $\mathbb{P}_{G_\sigma}$. Here, the CCA recognizes that Cholesky entries which correspond to non-zero entries in $\Omega$ can be regarded as functionally independent parameters in the Cholesky, whereas zeros in $\Omega$ which are not zeros in the Choleksy are actually functionally dependent parameters (even if they appear as non-zeros in the corresponding Cholesky). This insight follows from realizing that $\Omega$ and its Cholesky are simply two representations of the same quantity - hence the number of true functionally independent parameters in the Cholesky has to be the same as that of the $\Omega$ (i.e., the dimension of the parameter set cannot change because of a transformation). Thus, the entries in the Choleksy which correspond to zero entries in $\Omega$ can be regarded as functionally dependent. With this insight in hand, CCA modifies the entries of the Cholesky corresponding to these functional independent parameters (i.e., ``fill-in" entries), to obtain zeros in the corresponding positions at the level of $\Omega$. This process is essentially a projection of the Cholesky to the correct parameter space. Doing so enforces the desired sparsity pattern at the level of $\Omega$. This process of modification or adjustment of the Cholesky is facilitated by the fact that $\Omega$ is simply its Cholesky factor multiplied by the transpose of the Cholesky factor. In other words, the quadratic relationship between $\Omega$ and its Cholesky factor renders the adjustment of the fill-in entries a straightforward task (and importantly in closed form). In summary, note that the original graph $G$ has a given zero pattern  - some of which are lost by transferring the problem to the graph $G^D$. The additional edges introduced by going to $G^D$, and the sparsity that is lost because of this, is regained through adjusting the Cholesky parameter of $\hat{\Omega}^D$. The adjustment of the Cholesky to achieve the desired sparsity pattern at the level of $\Omega$ constitutes Step 2 of the CCA approach.

Altogether, the above closed form estimation and adjustment operations completely circumvent recourse to any iterative procedure. Specifics of the CCA methodology are now described through the aforementioned two step process.

\medskip

\noindent
\underline{\bf Step I:} The first step in our approach is to find an estimator of $\Omega$ in $\mathbb{P}_{G_\sigma^D}$. Recall that 
$G_\sigma^D$, which denotes the filled graph for $G_\sigma$, is a decomposable graph whose edge set subsumes 
the edge set of $G_\sigma$. When $\Omega \in \mathbb{P}_{G_\sigma^D}$ and if the underlying distribution is Gaussian,  
the MLE $\widehat{\Omega}^D$ for $\Omega$ is well-defined as long as $n$ is equal to or larger than the maximal clique size in $G_\sigma^D$, and can be obtained in closed form (see \cite{lrtzngphmd}). Let 
$\widehat{L}^D$ denote the Cholesky factor of $\widehat{\Omega}^D$. It can be shown (see for example 
\cite{AtayKayis:Massam:2005, Roverato:2002}) that the non-zero entries of the $j^{th}$ column of 
$\widehat{L}^D$ can be obtained as 
\begin{equation} \label{cholfactor}
\widehat{L}^{D, >}_{\cdot j} = -\frac{(S^{> j})^{-1} S^{>}_{\cdot j}}{\sqrt{S_{jj} - (S^{>}_{\cdot j})^T (S^{> j})^{-1} 
		S^{>}_{\cdot j}}} \mbox{ and } \widehat{L}_{jj} = \frac{1}{\sqrt{S_{jj} - (S^{>}_{\cdot j})^T (S^{> j})^{-1} 
		S^{>}_{\cdot j}}} 
\end{equation}

\noindent
for every $1 \leq j \leq p$. Here for any matrix $A$, $A^{> j} := ((A_{kl}))_{k,l > j, \; (k,j), (l,j) \in E_\sigma^D}$ and $A^{>}_{\cdot j} := (A_{kj})_{k > j, \; (k,j) \in E_\sigma^D}$. Note that 
$\widehat{\Omega}^D$ lies in $\mathbb{P}_{G_\sigma^D}$ and generally will not lie in the desired space $\mathbb{P}_{G_\sigma}$. We will refer to $\widehat{\Omega}^D$ and its Cholesky factor $\widehat{L}^D$ as 
{\it intermediate} estimators. Note that Step I produces estimators for $\Omega$-entries corresponding not only to edges in $E_\sigma$ but also for additional edges in $E^D_\sigma \setminus E_\sigma$. 

%
%

\medskip

\noindent
\underline{\bf Step II:} The second step in our approach is to modify or adjust the Cholesky factor $\widehat{L}^D$ to a lower triangular 
matrix $\widehat{L}$ such that $\widehat{\Omega} = \widehat{L} \widehat{L}^T \in \mathbb{P}_{G_\sigma}$, i.e., to eliminate the additional non-zero entries in our estimate of $\Omega$ that arise from working with a decomposable cover so that our estimator lives in the desired space. In particular, we construct $\widehat{L}$ by sequentially changing each fill-in entry of $\widehat{L}^D$ such that the corresponding entry of the 
concentration matrix (obtained by multiplying the resulting lower triangular matrix with its transpose) is zero. We start with 
the second row and move from entries with the lowest index to the highest index just before the diagonal. Whenever we 
encounter a fill-in entry along the way, say $(i,j)$, we set $\widehat{L}_{ij}$ as follows: 
\begin{equation} \label{cholupdate}
\widehat{L}_{ij} = - \frac{\sum_{k=1}^{j-1} \widehat{L}_{ik} \widehat{L}_{jk}}{\widehat{L}_{jj}}, 
\end{equation}

\noindent
thereby ensuring that the $(i,j)^{th}$ entry of the corresponding concentration matrix is zero. Note that the update in 
(\ref{cholupdate}) only uses entries from the $i^{th}$ and $j^{th}$ row. Also, any fill-in entries appearing on the right hand side 
in (\ref{cholupdate}) have already been updated at this point, and will not be changed subsequently. Hence, the resulting concentration matrix at the end of the second step of 
our algorithm lies in $\mathcal{P}_{G_\sigma}$ (see Lemma \ref{CCAproof} below for formal 
proof). The above construction by design also guarantees that $\widehat{L}_{ij} = 0 = 
\widehat{L}^D_{ij}$ if $i > j, (i,j) \notin E_\sigma^D$, and $\widehat{L}_{ij} = 
\widehat{L}^D_{ij}$ if $(i,j) \in E_\sigma$ or $i = j$, i.e., only the fill-in entries of $\widehat{L}^D$ are changed. 

We call the proposed statistical method Constrained Cholesky Approach 
({CCA}) and the correspondng algorithm the CCA algorithm - see Algorithm \ref{algorithm1} below. Note that $\widehat{\Omega}$ (the resulting CCA estimate) is obtained in a fixed and finite 
number of steps, and there is no repetition of an identical sequence of steps until convergence. The following 
lemma (proof in Supplemental Section C) establishes that $\widehat{\Omega}$ produced by CCA lies in 
$\mathbb{P}_{G_\sigma}$, i.e., it is positive definite and has the required pattern of zeros. 
\begin{lem} \label{CCAproof}
	Suppose $n$ is greater than the largest clique size in $G_\sigma^D$. Let $\widehat{\Omega}$ be the estimate of the 
	concentration matrix produced by CCA method (see also Algorithm \ref{algorithm1}). Then $\widehat{\Omega} \in \mathbb{P}_{G_\sigma}$. 
\end{lem}

\newtheorem{remark}{Remark}
\begin{remark}
The key idea used in the {CCA} algorithm is to first find a cover of the given graph for which a principled estimator can be easily computed (Step I), and then adjust the Cholesky values corresponding to the extra edges in the cover to obtain the desired sparsity in the precision matrix (Step II). This process works well for decomposable covers. A natural question to ask is whether there is some known class of graphs which would work even better, i.e., is the class of decomposable graphs maximal or optimal in some sense, or can we do better? We argue below that the proposed approach would either not work as effectively, or would even break down for other well-known classes of graphs studied in the graphical models literature. One candidate to consider is the class of homogeneous graphs (\cite{Letac:Massam:2007, Khare:Rajaratnam:2012}), which is a well-studied subclass of decomposable graphs. While the Gaussian MLE restricted to a homogeneous sparsity pattern is available in closed form, a homogeneous cover typically needs many more added edges compared to a decomposable cover. This would lead to many more fill-in entries, and hence increase the computational burden in Step II. Too many additional entries can also affect statistical precision. The same line of reasoning can be used to show that any special class of graphs which is a subset of the class of decomposable graphs (such as tress) would require additional edges, thus substantially increasing computational burden in Step II without any material benefit in Step I. On the other side, the class of Generalized Bartlett (GB) graphs (\cite{KRS:2018}) contains the class of decomposable graphs as a special case. Unlike decomposable graphs, the Gaussian MLE for general GB graphs is not available in closed form. Hence, an iterative algorithm would be needed for Step I, which in turn would significantly increase computational burden. So, any gains and justification thereof for using a Generalized Bartlett cover would have to come from Step II, since there are fewer extra zeros to correct in the intermediate estimator of $\Omega$. However, these savings cannot reverse the substantial additional computational cost required for Step I. As the class of decomposable graphs is the maximal class of graphs for which the Gaussian MLE and its Cholesky factor can be obtained in closed form, any larger class of graphs (such as the GB class) would not be amenable to the proposed CCA approach. The above arguments underscore the correct balance that decomposable graphs achieve.
\end{remark}

\noindent
{\color{black} Another question is whether the {CCA} estimators $\widehat{\Omega}$ and 
$\widehat{L}$ can be obtained as solutions to an optimization problem. 
Note that by definition the intermediate estimator $\widehat{\Omega}^D$ in Step I above minimizes the negative Gaussian log-likelihood (subject to 
$\Omega \in \mathbb{P}_{G^D_\sigma}$). Using the decomposability of 
$G^D_\sigma$, the intermediate Cholesky estimator $\widehat{L}^D$ can thus be 
expressed as 
\begin{equation} \label{interoptim}
\widehat{L}^D = \mbox{argmin}_{L \in \mathbb{L}_{G^D_\sigma}} \left\{ 
tr(LL^T S) - 2 \log |L| \right\}. 
\end{equation}

\noindent
The next lemma (proof in Supplemental Section D) identifies the {CCA} estimator 
$\hat{L}$ obtained in Step II above as the unique solution to an optimization problem 
(given the intermediate estimator $\widehat{L}^D$ obtained in Step I). 
\begin{lem} \label{step2}
Given the intermediate estimator $\widehat{L}^D$, the {CCA} estimator 
$\widehat{L}$ uniquely minimizes (over the set $\mathbb{L}_{G_\sigma}$) 
the function 
$
h(L) = \sum_{(i,j) \in E_\sigma \mbox{ or } i=j} (\widehat{L}^D_{ij} - 
L_{ij})^2. 
$
\end{lem}}

\noindent
Since the intermediate estimator $\widehat{L}^D$ is itself a solution to an optimization 
problem, the above lemma can be combined with (\ref{interoptim}) to obtain the following 
representation of the final {CCA} estimator. 
\begin{lem}
Given the sample covariance matrix $S$, the ordered graph $G_\sigma$ and its filled graph $G^D_\sigma$, the final CCA estimator $\widehat{L}$ can be expressed as 
$$
\widehat{L} = \mbox{argmin}_{L \in \mathbb{L}_{G_\sigma}} \sum_{(i,j) \in E_\sigma \mbox{ or } i=j} \left( \left( \mbox{argmin}_{L \in \mathbb{L}_{G^D_\sigma}} \left\{ 
tr(LL^T S) - 2 \log |L| \right\} \right)_{ij} - L_{ij} \right)^2. 
$$
\end{lem}

\noindent
The above lemma shows how CCA avoids an iterative optimization approach by decomposing the estimation problem into two steps. The first step projects the problem to a relevant space involving chordal graphs, and has a closed form solution. The second step projects this solution on to the desired space thus ensuring the correct sparsity pattern and preserving positive definiteness, and requires a single pass through the fill-in entries of $\widehat{L}^D$. These two steps together circumvent iterative/repetitive high-dimensional computations: a single iteration is all that is required at each of the two steps of CCA. 

\begin{remark} \label{non-parametric}
    \textcolor{black}{(Non-parametric interpretation of CCA) We now note that the intermediate estimator $\widehat{L}^D$ can in fact be seen as a non-parametric estimator which does not rely on the Gaussian functional form. In particular, the corresponding precision matrix estimator $\widehat{\Omega}^D = \widehat{L}^D (\widehat{L}^D)^T$ is the unique matrix in $\mathbb{P}_{G^D_\sigma}$ which satisfies the constraint that the $(i,j)^{th}$ entry of its inverse equals $S_{ij}$ (the $(i,j)^{th}$ entry of the sample covariance matrix) for 
    every $(i,j) \in E^D_\sigma$. To see this more formally, as in \cite{Letac:Massam:2007}, let $I_{G^D_\sigma}$ be the linear space of symmetric incomplete matrices $A$ with the $(i,j)^{th}$-entry missing for every $(i,j) \notin E$. Let $Q_{G^D_\sigma}$ be the collection of all $A \in I_G$ such that any submatrix of $A$ corresponding to a maximal clique of $G^D_\sigma$ is positive definite. For any $A \in Q_{G^D_\sigma}$, the decomposability of $G^D_\sigma$ along with a result in \cite{GRONE1984109} implies the existence of a unique completion $\tilde{A}$ of $A$ such that $\tilde{A}^{-1} \in \mathbb{P}_G$. Now, let $S^D$ denote the matrix in $I_{G^D_\sigma}$ obtained by removing all entries with indices outside $E^D_\sigma$. If $n$ is larger than the largest clique size in $G^D_\sigma$, then it follows that $S^D \in Q_G$. It now follows by \cite[Theorem 2.1]{Letac:Massam:2007} that $\widehat{\Omega}^D = (\tilde{S}^D)^{-1}$. Note specifically that Step 1 on the CCA approach can be implemented without invoking the Gaussian functional form. As Lemma \ref{step2} demonstrates, the second step of the CCA approach (constructing the CCA estimator $\hat{L}$ for $\hat{L}^D$) is again non-parametric/distribution-free in nature and only adjusts the fill-in entries of $\widehat{L}^D$ to ensure that the CCA precision estimator lies in $P_{G_\sigma}$. Given the non-parametric flavor of the entire CCA procedure, the CCA estimator is more robust. It is thus more immune to outliers compared to the IPF and its variants which in turn fundamentally depend on the Gaussian likelihood.}
\end{remark}

\begin{algorithm}[h]
	\caption{The CCA Algorithm}
	\label{algorithm1}
	\begin{flushleft}
	\vspace{0.1in}
	INPUT - {Sample covariance matrix $S$}\\
	INPUT - {Sparsity pattern encoded in a graph $G$}\\
 
	Compute RCM fill reducing ordering $\sigma$\\
	Compute filled graph $G_\sigma^D$\\
	Compute Cholesky factor $\widehat{L}^D$ of $\widehat{\Omega}^D$ (as specified in (\ref{cholfactor}))\\
	Set $\widehat{L} \leftarrow \widehat{L}^D$
    \end{flushleft}

    \begin{algorithmic}
	\ForAll {$i = 2$ to $p$}
	\ForAll {$j = 1$ to $i-1$}
	\If {$(i,j) \notin E_\sigma$} 
    \State Set $\widehat{L}_{ij} = 
	\frac{-\sum_{k=1}^{j-1} \widehat{L}_{ik} \widehat{L}_{jk}}{\widehat{L}_{jj}}$
    \EndIf
    \EndFor
    \EndFor
    \end{algorithmic}

    \begin{flushleft}
    OUTPUT - {Return final estimates $\widehat{L}$ and $\widehat{\Omega} = \widehat{L} \widehat{L}^T$}
    \end{flushleft}
	\end{algorithm}

\subsection{Computational aspects of the {CCA} Algorithm} \label{sec:complexity}

\noindent
In this section, we provide relevant computational details and evaluate the computational complexity of each 
step in the {CCA} algorithm. (Algorithm \ref{algorithm1}). As part of 
calculating the computational complexity of the CCA algorithm, note the following. 
\begin{itemize}
	\item The complexity of the step to determine the RMC fill reducing ordering $\sigma$ is 
	$O(|E_\sigma|)$ (see \cite{Chan:George:1980}). 
	\item The complexity of computing $G_\sigma^D$ (the decomposable cover) using the elimination tree 
	method outlined in Section \ref{sec:cholfillin} is $O \left( \sum_{j=1}^p n_j^2 \right)$, where $n_j = \{i: 1 \leq j < i 
	\leq p, (i,j) \in E_\sigma^D\}$ for every $1 \leq j \leq p$. 
	\item Computational complexity of Step 1: The complexity of computing $\widehat{L}^D$ based on eq. (\ref{cholfactor}) is calculated as $O \left( p + \sum_{j=1}^p n_j^3 
	\right)$, since the complexity of computing the inverse of a matrix is the cube of the dimension of the matrix. 
	Of course, it only makes sense to use (\ref{cholfactor}) to compute $\widehat{L}^D$ if $G_\sigma^D$ is sparse. 
	For dense $G_\sigma^D$ we can use a direct Cholesky factorization to obtain $\widehat{L}^D$ with complexity $O(p^3)$. {\color{black} As a practical rule, we use direct computation of $\hat{L}^D$ when $|E_\sigma^D| > p^{5/3}$, since in this case by Jensen's inequality 
	\begin{equation} \label{eqjensen}
	\sum_{j=1}^p n_j^3 \geq p \left( \frac{1}{p} \sum_{j=1}^p n_j \right)^3 = \frac{|E_\sigma^D|^3}{p^2} > p^3. 
	\end{equation}
	 }
	\item Computational complexity of Step 2: The second step of obtaining $\widehat{L}$ from $\widehat{L}^D$ can be efficiently implemented by 
	observing the following. Note that $\widehat{\Omega} \in \mathbb{P}_{G_\sigma} \subseteq 
	\mathbb{P}_{G_\sigma^D}$. Since $G_\sigma^D$ is the filled graph for $G_\sigma$, it follows that 
	$\widehat{L}_{ij} = 0 = \widehat{L}^D_{ij}$ if $i > j, (i,j) \notin E_\sigma^D$. Also, by construction, 
	$\widehat{L}_{ij} = \widehat{L}^D_{ij}$ if $(i,j) \in E_\sigma$ or $i = j$. Hence, the only entries which will be 
	changed in the second step of Algorithm \ref{algorithm1} are the fill-in entries or the edges in $E_\sigma^D 
	\setminus E_\sigma$. The complexity of modifying these entries in the algorithm can be easliy calculated as $O \left( p \left| E_\sigma^D \setminus E_\sigma \right| \right)$. 
\end{itemize}

\noindent
Straightforward calculations show that the complexities in the first two bullets are dominated by those in the last two bullets above, leading to the following lemma. 
\begin{lem}[Computational complexity of CCA]
For any connected ordered graph $G_\sigma$, the computational complexity of the entire CCA algorithm is 
$$
O \left( \min \left( p + \sum_{j=1}^p n_j^3 + p \left| E_\sigma^D \setminus E_\sigma \right|, \; p^3 \right) \right). 
$$
\end{lem}

{We now quantify the computational complexity of state-of-the-art methods. Recall that 
Iterated Proportional Fitting (IPF) is the classical ``go-to method" for computing the MLE for $\Omega$ in a 
Gaussian concentration graphical model when the underlying sparsity pattern (encoded in the graph $G$) is known (\cite{speedkvrgr}). In particular, the goal is to compute 
\begin{equation} \label{mle}
\arg\min_{\Omega \in \mathbb{P}^+_{G}} tr(\Omega S) - \log det(\Omega). 
\end{equation}

\noindent
Note that each iteration of the IPF algorithm requires sequential conditional 
maximization of the Gaussian likelihood with respect to sub-matrices corresponding to 
maximal cliques of the given graph $G$. If a given maximal clique has $c$ vertices, then 
the corresponding conditional maximization involves inversion of a $(p-c)$-dimensional 
matrix and thus requires $O((p-c)^3)$ iterations. This can be expensive for sparse graphs 
where $c << p$ in general. Several modifications of the original IPF algorithm which 
increase computational efficiency in such settings have been proposed over the last two 
decades. The line of work in \cite{Hara:Takemura:2010,XGH:2011,XGT:2012} and others focuses on `local' 
updates of the clique based sub-matrices using triangulations and junction trees to avoid 
large matrix inversions. As pointed out in \cite{XGT:2015}, while these methods can 
significantly lower the amount of computations, they need significantly more memory. The 
{IPSP} method in \cite{XGT:2015} employs the strategy of partitioning the set of maximal 
cliques of $G$ into non-overlapping blocks, and then performing local updates for the 
clique sub-matrices in each block. All of the above methods are iterative and need an 
enumeration of the maximal cliques of the graph $G$. While such an enumeration can be done 
efficiently for some family of sparse graphs, it can also be very expensive in general and 
takes exponential time in the worst-case setting (see Example 5 below). In very recent work, \cite{Hogsgard:Lauritzen:2023} 
cleverly employ a version of Woodbury's identity in \cite{Harville:1977} to avoid enumeration of maximal 
cliques of $G$ as well as large matrix inversions in the conditional maximization steps of the IPF algorithm. 
We refer to this algorithm, based on the ``edgewise" and ``covariance-based" version of IPF developed in \cite{Hogsgard:Lauritzen:2023}, simply as {FIPF} 
(Fast IPF). 

An alternative approach to find the MLE in the Gaussian setting is described in \cite[Algorithm 17.1]{ESL} and 
implemented in the {\it glasso} package. Essentially, this approach uses a modification of the well-known Glasso algorithm (i.e., without the 
$\ell_1$ penalty term which accounts for the fact that the sparsity pattern in $\Omega$ is known). We will refer to this algorithm as the 
G-IPF algorithm. Note that similar to the IPF and its adaptations, {G-IPF}  repeats its iterations until an appropriate convergence 
threshold has been achieved. However, unlike these algorithms (except F-IPF), {G-IPF}  does not require an enumeration of the maximal cliques of 
$G$. A straightforward analysis of \cite[Algorithm 17.1]{ESL} and results in \cite{XGT:2015, Hogsgard:Lauritzen:2023} can be used to calculate the {\it overall} computational complexity of the {IPSP}, {FIPF} and {G-IPF}  algorithms. These complexities, along with the complexity of the proposed CCA algorithm (based on the discussion at the beginning of this subsection) are summarized in Table \ref{computational:complexity}. 
\begin{table}
\centering
\begin{tabular}{|c|c|}
\hline
Method & Overall complexity\\
\hline
     {CCA} & $O \left( \min \left( p + \sum_{j=1}^p n_j^3 + p \left| E_\sigma^D \setminus E_\sigma \right|, \; p^3 \right) 
	\right), 
	$ \\
    {G-IPF} & $T_{Gl} \times O \left( \sum_{j=1}^p \widetilde{n}_j^3 + p \left| E_\sigma \right| \right)$\\
    {FIPF} & $T_F \times O \left( p^2 |E_\sigma| \right)$\\
    {IPSP} & $T_{I} \times O \left( p^2 \sum_{U \in \mathcal{U}} |U| \right) + T_{enum}$\\
    \hline
\end{tabular}
\caption{Overall computational complexity of various undirected/concentration graph model estimators (for connected $G_\sigma$). For disconnected graphs, complexities can be computed over all the connected components and then added together. Here $T_{Gl}, T_F, T_I$ denote the number of repetitions needed to achieve the convergence threshold for {G-IPF}, {FIPF}, {IPSP} respectively, $T_{enum}$ is the time needed to enumerate the maximal cliques of $G$, $\widetilde{n}_j = \left| \{i: 1 \leq i \leq p, (i,j) \in E_\sigma\} \right|$, $n_j = \{i: 1 \leq j < i 
	\leq p, (i,j) \in E_\sigma^D\}$, $\mathcal{C}$ is the collection of all maximal cliques of $G$, and $\mathcal{U}$ is a partition of the $\mathcal{C}$.}
\label{computational:complexity}
\end{table}

The next lemma (proof in Supplemental Section E) compares the computational complexity of {CCA} with {FIPF} and {IPSP}. 
\begin{lem}(Computational Complexity: {CCA} vs. {FIPF} and {IPSP}) \label{lem_CF}
For an arbitrary graph $G$, the computational complexity of even one iteration of the {IPSP} algorithm or the FIPF algorithm is greater than or 
equal to that of the entire {CCA} algorithm. 
\end{lem}

\noindent
The next lemma (proof in Supplemental Section F) compares the computational complexity of {CCA} with {G-IPF} especially in the context when $p$ is large. To facilitate this analysis, we consider the class of $p$-vertex graphs with at least $p^{5/3}$ 
edges, denoted by $\mathcal{A}_p$. This is a large class of graphs with elements ranging from 
{\it moderately sparse graphs} (since $p^{5/3} = o \left( {{p}\choose{2}} \right)$), as well as {\it very dense graphs} (with more than 
$0.5 {{p}\choose{2}}$ edges). 
\begin{lem}(Computational Complexity: {CCA} vs. {G-IPF} ) \label{lem_CG}
Consider the class $\mathcal{A}_p$ of $p$-vertex graphs with more than $p^{5/3}$ edges. Then 
\begin{enumerate}[(a)]
\item For all graphs $G \in \mathcal{A}_p$, the {CCA} algorithm is computationally more efficient than even one iteration of the {G-IPF}  
algorithm. 
\item The class of $p$-vertex graphs belonging to $\mathcal{A}_p$ as a proportion of the total number of $p$-vertex graphs converges super-exponentially to $1$ as $p \rightarrow \infty$. In particular, 
$$
\frac{|\mathcal{A}_p|}{2^{{p}\choose{2}}} \geq 1 - \sqrt{\frac{2}{\pi}} \exp \left( - \left( 0.17 p^2 - 3p^{5/3} \log p \right) \right) \mbox{ for } p \geq 100.
$$
\end{enumerate}
\end{lem}

\noindent
In summary, compared to competing methods the computational complexity of the proposed CCA method is always lower for any dimension (when compared to FIPF and IPSP) or almost always lower (when compared to G-IPF) as the dimension $p$ gets larger, underscoring its direct relevance in modern settings. Equally or more importantly, this lower complexity is for a single iteration, bearing in mind that CCA requires only one iteration, while competing methods require many. These two properties together make a very strong and compelling case for the use of CCA in modern settings. 

We now turn to comparing the CCA and G-IPF approaches in the sparse graph setting (graphs with less than $p^{5/3}$ edges). Though a general comparison of the computational complexities of {CCA} and {G-IPF} is not available in this setting, important insights can be obtained through a case-by-case analysis. Recall that a single iteration of the {G-IPF}  algorithm has complexity of the order $\sum_{j=1}^p \widetilde{n}_j^3 + p \left| 
E_\sigma \right|$, while the single and {\it only} iteration of Algorithm \ref{algorithm1} has computational complexity of the order 
$\sum_{j=1}^p n_j^3 + p + p \left| E_\sigma^D \setminus E_\sigma \right|$ for adequately sparse graphs. {Since the 
G-IPF algorithm does not use the filled graph $G_\sigma^D$, comparing $\left| E_\sigma \right|$ and $\left| E_\sigma^D \setminus 
E_\sigma \right|$ (also $\tilde{n}_j$ and $n_j$) is not straightforward and hence general results comparing these quantities are not available. 
However, we are able to compare these quantities (and consequently the computational complexities of {CCA} vs. {G-IPF}, and those of {FIPF} and {IPSP}) 
in some key examples in the next subsection.}}

\subsection{Examples and illustrations} \label{sec:example:classes}

\noindent
{\it Example 1 (grid)} An $a \times b$ grid is a common sparse non-decomposable graph with 
$p = (a+1)(b+1)$ vertices. {For the default column-based ordering $\sigma$ (depicted in Figure \ref{figgrid1} (middle) with $a=b=3$) the filled graph $G_\sigma^D$ 
can be obtained by adding a NE-SW diagonal in each of the $ab$ unit squares in the grid. The RCM ordering (Figure \ref{figgrid1}, right) again leads to the same NE-SW diagonals as fill-in edges. For the default row-based ordering $\sigma$ (depicted in Figure \ref{figgrid2} (left) with 
$a=b=3$) the filled graph $G_\sigma^D$ can be obtained by adding a NW-SE diagonal in each of the $ab$ unit squares in the grid. It is important to note that the orientation of these additional diagonals depends on the imposed ordering $\sigma$. For example, if we add diagonals with a NW-SE orientation, then the resulting graph does not admit the column-based ordering in Figure \ref{figgrid1} (middle) as a perfect vertex elimination scheme. In general, adding diagonals with different orientations in each unit square of the grid may not lead to a decomposable graph (see 
Figure \ref{figgrid2} (right) for an example).  An example where adding diagonals of different orientations leads to a decomposable graph is provided in Figure \ref{figgrid2} (middle), 
and a corresponding perfect vertex elimination scheme is also provided. 
\begin{figure}
\centering
\includegraphics[width=5in,height=3in]{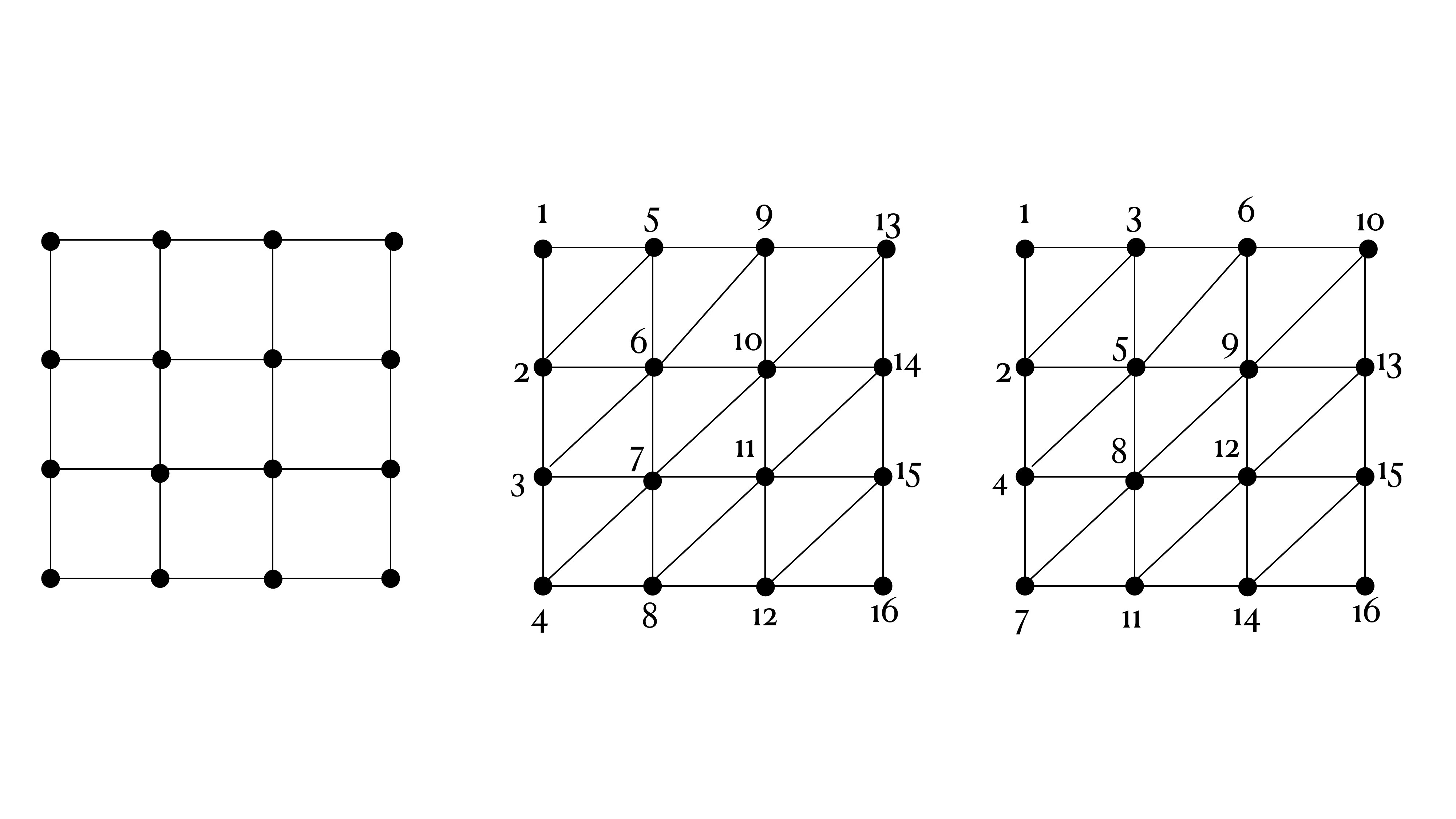}
\vspace{-0.5in}
\caption{(left) A $3 \times 3$ grid with $16$ vertices. (middle) Filled graph for column-based ordering of vertices with added diagonals in 
NE-SW orientation. (right) Filled graph for RCM ordering.}
\label{figgrid1}
\end{figure}

\begin{figure}
\centering
\includegraphics[width=5in,height=3in]{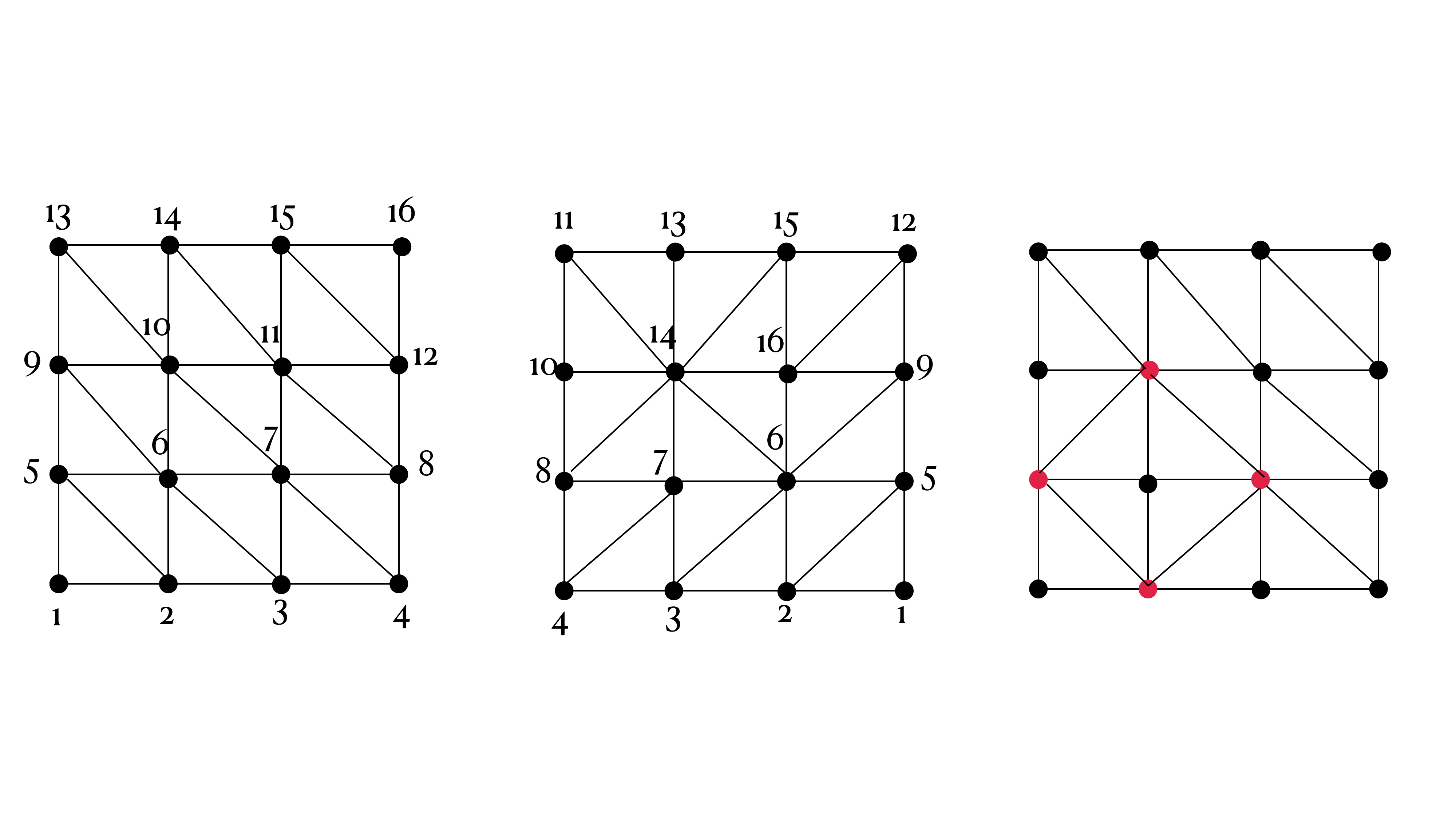}
\vspace{-0.5in}
\caption{(left) Filled graph for row-based ordering of vertices with added diagonals in NW-SE orientation. (middle) An example when 
adding diagonals of different orientations leads to a decomposable graph, and the corresponding perfect vertex elimination ordering.  
(right) An example when adding diagonals of different orientations does not lead to a decomposable graph, as the induced subgraph 
on the highlighted vertices is a $4$-cycle.}
\label{figgrid2}
\end{figure}

Let $\sigma$ be the RCM ordering depicted in Figure \ref{figgrid1} (right). In this case, 
straightforward computations show that 
$$
\left| E_\sigma^D \setminus E_\sigma \right| = ab < 2ab+a+b = \left| 
E_\sigma \right|
$$ 

\noindent
and 
$$
\sum_{j=1}^p n_j^3 = 27ab+a-11b < 64ab - 10a-10b-12 = \sum_{j=1}^p 
\tilde{n}_j^3. 
$$

\noindent
In this case the {CCA} algorithm and a single iteration of the {G-IPF}  algorithm both have complexity $O(p^2)$. 
Also, the underlying graph has $O(p)$ maximal cliques of size $2$, and it follows from Table \ref{computational:complexity} that 
both {FIPF} and {IPSP} have computational complexity $O(p^3)$ in this case. 

With the above comparisons in mind, in modern environmental and earth science applications, there is a compelling need to model covariates on a spatial grid and their corresponding dependencies in a sparse manner. In particular, the sparse modeling of partial correlations (and conditional independences) on a grid has been shown to be highly effective in modern climate and paleo-climate applications. In such applications the dimension can be much larger than the available sample size (see \cite{GRE:2015}). Here the goal is to not just estimate a sparse Markov Random Field, but it is 
critically important to also quantify the uncertainty inherent to such estimation, through resampling or other methods (see \cite{GRE:2015}). When estimation is not in closed form, quantifying uncertainty through multiple bootstrap estimations of the inverse covariance matrix can be extremely computationally prohibitive. The grid example above thus demonstrates direct applicability of the proposed CCA method in a contemporary application domain.

\medskip

\noindent
{\it Example 2 (cycle)}. The $p$-cycle is another standard example of a sparse non-decomposable graph. A simple 
induction argument shows that $\left| E_\sigma^D \setminus E_\sigma \right| = p-3 < p = \left| E_\sigma \right|$ if $\sigma$ is chosen 
to be the RCM fill reducing ordering of the vertices (see Figure 
\ref{cycle}). In this case, $\tilde{n}_j = n_j = 2$ for every $1 \leq j \leq p-1$, $\tilde{n}_p = 2$ and $n_p = 0$. 
Hence, $\sum_{j=1}^p n_j^3 < \sum_{j=1}^p \tilde{n}_j^3$. { Hence, both the {CCA} and one iteration of {G-IPF}  have 
complexity $O(p^2)$ in this case. When considering the competing methods of FIPF and IPSP, note that since a $p$-cycle has $p$ maximal cliques of size $2$, it follows from 
Table \ref{computational:complexity} that both these methods have per iteration computational complexity $O(p^3)$ in this case. 
The IHT algorithm in \cite{XGH:2011} has a complexity of $O(p)$ (per iteration) in this case. However, as mentioned in \cite{XGT:2015}, this algorithm can be prohibitively memory intensive in general.} 
\begin{figure}
\centering
\includegraphics[width=5in,height=2.5in]{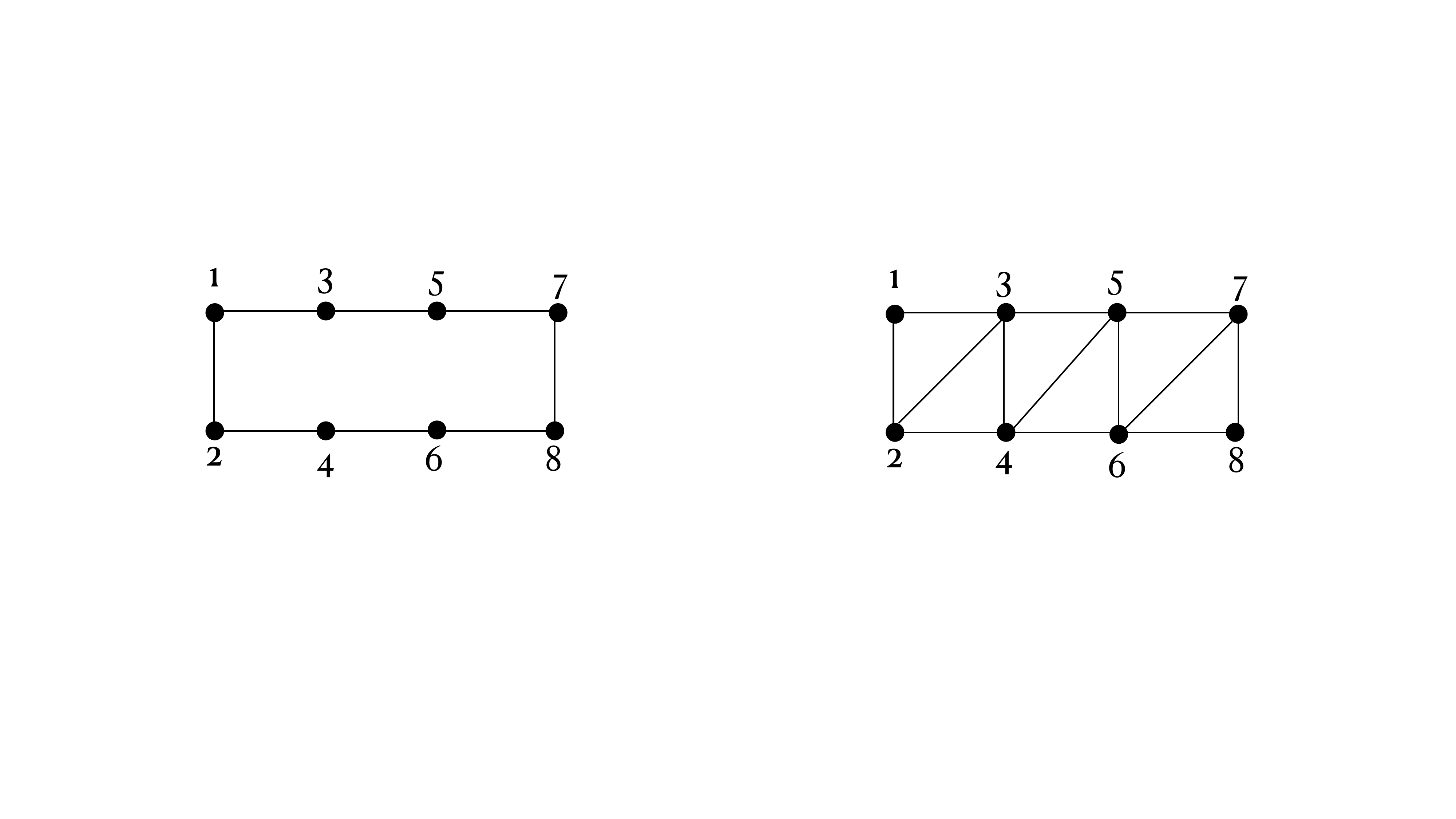}
\vspace{-0.9in}
\caption{An $8$-cycle with RCM ordering of vertices (left), 
corresponding filled graph (right).}
\label{cycle}
\end{figure}

\medskip

\noindent
{\it Example 3 (almost complete graph)}. { The largest/densest non-decomposable graph with $p$ vertices can be obtained 
by taking a 
complete graph (with $\sigma$ being the identity permutation), and removing the edges $(p, p-2)$ and $(p-1, p-3)$.} Then the induced 
subgraph on the subset of vertices $\{p,p-1,p-2,p-3\}$ is a $4$-cycle, which implies $E_\sigma$ is not a decomposable graph. In this 
case, the filled graph $E_\sigma^D$ can be obtained by adding the edge $(p,p-2)$ to $E_\sigma$. In this dense case, the 
computational complexity of CCA, based on the discussion corresponding to eq. (\ref{eqjensen}), is $O(p^3)$. On the other hand for G-IPF, 
$$
\sum_{j=1}^p {n_j}^3 = \sum_{k=1}^{p-1} k^3 - (3^3 - 2^3) = \frac{(p-1)^2 p^2}{4} - 19, 
$$

\noindent
{which implies that the computational complexity of one iteration of {G-IPF}  is $O(p^4)$: an entire order of magnitude larger. When considering FIPF and IPSP, note that the underlying graph has four maximal cliques of size $p-2$. Hence, the computational complexity 
of both {FIPF} and {IPSP} (per iteration) is $O(p^3)$ by Table \ref{computational:complexity}.} 

\medskip

\noindent
{ {\it Example 4 (Bipartite graph)}. Consider a bipartite 
graph setting, where $V = A \uplus B$ with $|A| = m$ and $|B| = p-m$, and 
$(u,v) \in E$ if and only if $u \in A, v \in B$ or $u \in B, v \in A$. We assume without loss of generality that $m \leq p/2$. For this non-decomposable graph, consider any vertex ordering $\sigma$ which assigns labels $\{p,p-1, \cdots, p-m+1\}$ to vertices in $A$, and the remaining labels to vertices in $B$. Such an ordering can be converted to an RCM ordering with just one change to the vertex labels, but we consider the above partition based ordering for simplicity of exposition. In this case, $G_\sigma^D$ can be obtained by adding edges so that each vertex in A shares an edge with all other vertices in $A$. In particular, $|E_\sigma^D \setminus E_\sigma| = {{m}\choose{2}}$. It follows that $n_j = m$ for $1 \leq j \leq p-m$, and $n_j = p-j$ for $j > m$. On the other hand $\tilde{n}_j = p-m$ for $1 \leq j \leq m$ and $\tilde{n}_j = m$ for $j > m$. Consider the `sparse' case, when $m = o(p^{2/3})$ and $\hat{L}^D$ is computed using (\ref{cholfactor}). In this case, note that 
$$
p + \sum_{j=1}^p n_j^3 + p|E_\sigma^D \setminus E_\sigma| = (p-m)m^3 + \sum_{j=1}^m j^3 + p + pm^2 \leq C pm^3 
$$

\noindent
Since $p/2 < p -m < p$, it follows that the computational complexity of 
CCA is $O(pm^3)$. Similarly, note that 
$$
\sum_{j=1}^p \tilde{n}_j^3 + p|E_\sigma| = (p-m)m^3 + m(p-m)^3+pm(p-m)
$$

\noindent
which implies that the computational complexity of {G-IPF}  (per iteration) 
is $O(mp^3)$. Now let us consider the other competing methods FIPF and IPSP. Since there are $m(p-m)$ maximal 
cliques, each of size $2$, it follows that the computational complexity of {FIPF} and 
{IPSP} is $O(mp^3)$ as well. Since $m = o(p^{2/3})$, we conclude that the 
computational complexity of the entire {CCA} algorithm is an order of 
magnitude smaller than the per iteration complexity of the other three 
algorithms. In the `dense' case where $m$ is at least $p^{2/3}$, and 
$\hat{L}^D$ is computed using the usual Cholesky factorization approach, 
the complexity of {CCA} is $O(p^3)$, while the complexity for the competing methods is 
at least $O(p^{3+2/3})$. 

\medskip

\noindent
{\it Example 5 (Multipartite graph - exponential number of cliques)}. Consider a setting, where $p = 3m$, 
$V = \uplus_{i=1}^m A_i$ with each $A_i$ having $3$ vertices, and $(u,v) 
\in E$ if and only if $u \in A_k, v \in A_{k'}$ for $k \neq k'$. Consider 
any ordering $\sigma$ which assigns labels $\{3i-2, 3i-1, 3i\}$ to 
vertices in $A_i$ for $1 \leq i \leq m$. This is a dense graph with 
${{p}\choose{2}}-p$ edges, and the computational complexity of {CCA} is 
$O(p^3)$. Since $\tilde{n}_j = p-2$ for every $j$, it follows that the computational complexity of {G-IPF}  (per iteration) is $O(p^4)$. With regards to the other two methods FIPF and IPSP, there are an exponential number of maximal cliques ($3^{p/3}$ in particular), and hence the computational complexity of these methods (per iteration) is exponential in $p$. 

\medskip

\noindent
For comparison purposes, the results in the examples above are summarized in 
Table \ref{tabexample}. The analysis above facilitates a detailed understanding of the properties of the proposed CCA method. In particular, the computational advantages of the proposed {CCA} algorithm stem from three properties: (a) {G-IPF} , {FIPF} and {IPSP} have to repeat their  
iterations $T$ times, while no such repetition is necessary for the CCA method, (b) {FIPF} and {IPSP} (not {G-IPF}  though) require an enumeration of the maximal cliques of the graph $G$, which can be expensive, especially for moderately sparse or dense graphs, and (c) in a large majority of settings, even a single iteration of the {CCA} algorithm is in general computationally more efficient than even a 
single iteration of the {G-IPF}  and other algorithms. Together, these three properties contribute to the significant computational advantage of the CCA approach.}

\begin{table}
\centering
	\begin{tabular}{|c|c||c|c|c|c|}
		\hline
  Non-decomp. graph & Sparsity & CCA & G-IPF & FIPF & IPSP\\
  \hline \hline
  Grid & Sparse & ${\bf O(p^2)}$ & $T_G \times O(p^2)$  & $T_F \times O(p^3)$ & $T_I \times O(p^3)$\\
  \hline
  Cycle & Sparse & ${\bf O(p^2)}$ & $T_G \times O(p^2)$  & $T_F \times O(p^3)$ & $T_I \times O(p^3)$\\
  \hline
  Bipartite & Sparse & ${\bf O(pm^3)}$ & $T_G \times O(mp^3)$ & $T_F \times O(pm^3)$ & $T_I \times O(pm^3)$\\
  $(m,p-m)$ partition & Dense & ${\bf O(p^3)}$ & $T_G \times \Omega(p^{11/3})$ & $T_F \times \Omega(p^{11/3})$ & $T_I \times \Omega(p^{11/3})$\\
   \hline
   Almost complete & Dense & ${\bf O(p^3)}$ & $T_G \times O(p^4)$  & $T_F \times O(p^3)$ & $T_I \times O(p^3)$\\
   \hline
   Multipartite & Dense & ${\bf O(p^3)}$ & $T_G \times O(p^4)$  & $T_F \times e^{O(p)}$ & $T_I \times e^{O(p)}$\\
   exponential \# cliques  & & & & &\\
   \hline
  \end{tabular}
  \caption{Computational complexity comparison of CCA with other competing methods for some key non-decomposable graphs. For iterative methods, the complexity of a single iteration is multiplied with the total number of iterations ($T_G$ for G-IPF, $T_F$ for FIPF, $T_I$ for IPSP). The lowest computational complexity for each row is highlighted in bold.}
  \label{tabexample}
\end{table}


\section{High-dimensional asymptotic properties} \label{sec:asymptotics}

\noindent
{ In this section we examine the large sample properties of the proposed estimators in a high-dimensional setting. Consider once more the setting in Section \ref{sec:congraphmd}. We will allow the number of variables $p = p_n$ to increase with the sample size $n$. We consider a true data generating model under which i.i.d. observations ${\bf Y}_{n,1}, \cdots, {\bf Y}_{n,n} \in \mathbb{R}^{p_n}$ are generated from a distribution with mean ${\bf 0}$ and precision matrix $\bar{\Omega}_n \in \mathbb{P}^+_{p_n}$ for every $n \geq 1$. In other words, $\{\bar{\Omega}_n\}_{n \geq 1}$ denote the sequence of true inverse covariance matrices. Let  $\{\bar{L}_n\}_{n \geq 1}$ denote the corresponding sequence of Cholesky factors of $\bar{\Omega}_n$, and let $\{\bar{\Sigma}_n\}_{n \geq 1}$ denote the corresponding sequence of true covariance matrices. The graph $G = G_n$ encodes the sparsity pattern in $\bar{\Omega}_n$, $\sigma = \sigma_n$ denotes a given fill-reducing ordering for $G$, and $G^D_\sigma$ denotes the corresponding filled graph. Let $\bar{P}$ denote the probability measure 
underlying the true data generating model described above. 

We now introduce some quantities related to $G_\sigma$ and $G^D_\sigma$ 
which will be needed for specifying the convergence rate of the {CCA} based 
estimators. Let $a_n^D$ denote the product of the maximum number of 
non-zero entries in any column of $\bar{L}_n$ and the maximum number of 
non-zero entries in any row of $\bar{L}_n$. In particular 
$$
a_n^D = \left( 1 + \max_{1 \leq j \leq p-1} |\{u > j: (u,j) \in 
E^D_\sigma\}| \right) \times \left( 1 + \max_{2 \leq i \leq p} |\{u < i: 
(i,u) \in E^D_\sigma\}| \right). 
$$

\noindent
The quantity $\tilde{a}^D_n$ is similarly defined below, although focusing 
exclusively on the fill-in entries only as opposed to the entire filled graph. 
In particular 
$$
\tilde{a}_n^D = \left( \max_{1 \leq j \leq p-1} |\{u > j: (u,j) \in 
E^D_\sigma \setminus E_\sigma\}| \right) \times \left( \max_{2 \leq i 
\leq p} |\{u < i: (i,u) \in E^D_\sigma \setminus E_\sigma\}| \right). 
$$

\noindent
Let $c = |E^D_\sigma \setminus E_\sigma|$ be the number of fill-in entries. These fill-in entries can be 
ordered based on the traversal path described in Step II of the {CCA} algorithm. For every $1 \leq 
r \leq c$, let $\mathcal{F}^D_r$ denote the collection of previous 
fill-in entries which contribute to the computation of the $r^{th}$ fill-in 
entry in Algorithm \ref{algorithm1}, see RHS of (\ref{cholupdate}). In 
particular if the $r^{th}$ fill-in entry is $(i,j)$, then 
\begin{eqnarray*}
\mathcal{F}^D_r
&=& \left\{\tilde{r}: 1 \leq \tilde{r} \leq r, \exists 
k < j < i \mbox{ such that } \tilde{r}^{th} \mbox{ fill-in entry is }
(j,k) \mbox{ with } (i,k) \in E^D_\sigma \right\}\\
& & \uplus \left\{\tilde{r}: 1 \leq \tilde{r} \leq r, \exists 
k < j < i \mbox{ such that } \tilde{r}^{th} \mbox{ fill-in entry is }
(i,k) \mbox{ with } (j,k) \in E^D_\sigma \right\}. 
\end{eqnarray*}

\noindent
We now state two of the three assumptions that are needed for our asymptotic results. Both of these 
assumptions are quite standard in the high-dimensional asymptotic literature. 
\begin{itemize}
	\item (A1 - Bounded Eigenvalues) There exists $\delta>0$ (not depending on $n$) such that 
	$
	0 < \delta \leq \lambda_{\min} (\bar{\Omega}) \leq \lambda_{\max} (\bar{\Omega}) \leq {\delta}^{-1} < \infty. 
	$ 
    \item (A2 - Sub Gaussianity) The random vectors ${\bf Y}_{n,1},\dots, 
    {\bf Y}_{n,n}$ are sub-Gaussian i.e., there exists a constant 
    $c > 0$ such that for every ${\bf x} \in \mathbb{R}^p$, $E \left[ 
    e^{{\bf x}' {\bf Y}^i} \right] \leq e^{c {\bf x}' {\bf x}}$. Along with Assumption A1, this in particular implies that the sub-Gaussian norm of ${\boldsymbol \alpha}^T {\bf Y}^i$, for any ${\boldsymbol \alpha}$ with ${\boldsymbol \alpha}^T {\boldsymbol \alpha} = 1$, is uniformly bounded by a constant $\kappa$. 
\end{itemize}

\noindent
\noindent
Assumption A1 states that the eigenvalues of the sequence of true covariance matrices are uniformly 
bounded. As mentioned above, these are standard assumptions in high-dimensional covariance asymptotics, 
see \cite{bckllvncrt, RBLZ:2008, PWZZ:2009, xiangkhareghosh}. Before stating the third required 
assumption, we define a function $g^D$ which maps $\{1,2, \cdots, c\}$ to $\mathbb{R}_+$ recursively as 
follows. 
$$
g^D (1) = \frac{6}{\delta} \mbox{ and } g^D (r) = \frac{3}{\delta}
\sum_{\tilde{r} \in \mathcal{F}^D_r} g^D (\tilde{r}) 
$$

\noindent
for $2 \leq r \leq c$. This function is critical in capturing the 
propagation of errors while modifying the fill-in entries to compute the 
CCA Cholesky estimate $\hat{L}$ from $\hat{L}^D$. The third required 
assumption stipulates that $p$ increases with $n$ at an appropriate sub-exponential rate depending on the graph based quantities defined above. 
\begin{itemize}
	\item (A3 - Growth rate for $p_n$) \hspace{0.2in} $\frac{s_{CCA} \log p}{n} \to 0$ as $n \to \infty$, 
     where 
     $$
     s_{CCA} = \min \left( a^D_n, |E^D_\sigma|+1 \right) \tilde{a}^D_n 
     \left( 1 + \max_{1 \leq r \leq c} g^D (r) \right)^2. 
     $$ 
\end{itemize}

\noindent
With the required assumptions and notation in hand, we establish our consistency result for {CCA}. Part 
(a) of the result establishes bounds for the intermediate estimator $\hat{L}^D$, and part (b) established 
bounds for the final {CCA} estimators $\hat{L}$ and $\hat{\Omega}$. 
\begin{thm}(Convergence rates for CCA) \label{CCA_convergence_rate}
Suppose that assumptions A1-A3 are satisfied. Then the following results hold. 
	\begin{enumerate}[(a)]
		\item There exists a constant $K^*$ such that 
		$$
		\bar{P} \left( \norm{\widehat{L}^D - \bar{L}}_2 \leq K^* \sqrt{\frac{\min \left( a^D_n, |E^D_\sigma|+1 \right) \log p}{n}} 
  \right) \rightarrow 1 \mbox{ as } n \to \infty. 
		$$
		
		\item There exist constants $K$ and $K'$ such that 
		$$
		\bar{P} \left( \left\{\norm{\widehat{L} - \bar{L}}_2 \leq K \sqrt{\frac{s_{CCA} \log p}{n}} \right\} \cap \left\{ \norm{\widehat{\Omega} - \bar{\Omega}}_2 
        \leq K' \sqrt{\frac{s_{CCA} \log p}{n}} \right\} \right) \rightarrow 1 \mbox{ as } n \to \infty. 
		$$
	\end{enumerate}
\end{thm}

\noindent
Like IPF, our work/model assumes that the graph $G$ is known. When it is not known, it can be constructed 
using thresholding methods for sparse inverse covariance estimates as in \cite{Hero:Rajaratnam:2012} or \cite{zhang2021quadratic}.

\begin{remark} (Weak convergence in the fixed $p$ setting)
When $p$ does not vary with $n$, distributional convergence results which 
show that $\widehat{\Omega}$ is $\sqrt{n}$-consistent for $\bar{\Omega}$ 
can be established under mild moment assumptions. In particular, assume 
that the underlying distribution $F$ of the i.i.d. observations ${\bf Y}_1, {\bf Y}_2, \cdots, {\bf Y}_n$ satisfies the assumption 
$\int_{\mathbb{R}^p} \prod_{i=1}^p |x_i|^{a_i} dF({\bf x}) < \infty$, 
where $a_1, a_2, \cdots, a_p$ are non-negative integers which satisfy $\sum_{i=1}^p a_i = 4$. Under this assumption, it can be shown that $\sqrt{n}(\widehat{\Omega} - \bar{\Omega})$ converges to a multivariate Gaussian limit as $n \rightarrow \infty$ 
(see Supplemental Section B).
\end{remark}

\subsection{Comparison with operator norm convergence rate for graphical Gaussian MLE} \label{mlebound}

\noindent
High-dimensional convergence rates for the graphical Gaussian MLE (solution of (\ref{mle})) are not 
directly available in the current literature. However, minor modifications to the argument in 
\cite[Theorem 1]{RBLZ:2008} lead to a Frobenius norm convergence rate of 
$\sqrt{\frac{(1+|E_\sigma|)\log p}{n}}$ for the graphical Gaussian MLE, 
under Assumptions A1-A2 (and a modified Assumption A3 which states 
that $\sqrt{\frac{(1+|E_\sigma|) \log p}{n}} \rightarrow 0$ as $n \rightarrow 
\infty$). Since the Frobenius norm dominates the operator norm, the above rate of $\sqrt{\frac{(1+|E_\sigma|) \log p}{n}}$ 
also applies to the operator norm. To our knowledge, this is the best available operator norm convergence rate for 
the constrained Gaussian MLE (with a general graph $G$). Hence, the comparison between the available operator norm convergence rates for the {CCA} and the graphical Gaussian MLE 
reduces to the comparison between the order of the quantities {\it $s_{CCA}$} and 
$|E_\sigma|$. 

Note that {\it $s_{CCA}$ is a product of two factors}: $\min(a^D_n, |E_\sigma^D|+1)$ (corresponding 
to Step I of Algorithm \ref{algorithm1}) and $\tilde{a}^D_n 
\left( 1 + \max_{1 \leq r \leq c} g^D (r) \right)^2$ (corresponding to 
Step II of Algorithm \ref{algorithm1}). We are able to obtain the term $\min(a^D_n, |E_\sigma^D|+1)$ (instead of just $|E^D_\sigma|+1$) in the operator norm bounds because the intermediate 
estimators $\hat{L}^D$ and $\hat{\Omega}^D$ are available in closed form. 
While $|E^D_\sigma| > |E_\sigma|$ for all ordered graphs, in many 
settings $a^D_n$, the product of the maximum number of non-zero entries 
in any row of $\bar{L}$ and the maximum number of non-zero entries in any 
column of $\bar{L}$, is of a lower order than $|E_\sigma|$. Hence, if the 
multiplicative factor corresponding to Step II does not grow too fast, 
$s_{CCA}$ can be of a lower order than $|E_\sigma|$. {\color{black} To summarize, key properties of the CCA approach – (i) the availability of the intermediate estimator in closed form, and (ii) the final estimator requiring a finite number of tractable algebraic changes to the intermediate estimator – together allow us to directly analyze the operator norm of the estimation error and establish corresponding rates. Importantly, we have a high-dimensional setting where better computational properties of the CCA approach translate to, and are directly relevant for, establishing better statistical properties. In this sense, these two properties are complementary. Note that the graphical Gaussian MLE however does not readily allow convergence in the operator norm, forcing existing analyses to target the Frobenius norm of the estimation error. The corresponding rates for the Frobenius norm are then borrowed for the operator norm, leading to looser bounds in high-dimensional settings.}

While a general comparison of {\it $s_{CCA}$} and $E_\sigma$ is not 
available, we undertake a comparison in some of the key examples 
considered at the end of Section \ref{sec:complexity}. It turns out that 
in most of these examples, {\it $s_{CCA}$} is of a smaller (or same) order as 
$|E_\sigma|$. We now consider the five classes of graphs from Section \ref{sec:example:classes}, and compare the available statistical accuracy rates 
of the CCA algorithm with that of the graphical Gaussian MLE for each of these classes. 



\medskip

\noindent
{\it Example 1 (grid) contd.} Consider an $a \times b$ grid with $p=(a+1)
(b+1)$ vertices. If $\sigma$ is the RCM fill reducing ordering 
(see Figure \ref{figgrid1}, right), then the maximum number of non-zero entries in any row or column of $\bar{L}$ 
is $4$, which implies that $\min(a^D_n, |E^D_\sigma|) = O(1) << 2ab+a+b = 
|E_\sigma|$. In this case $\mathcal{F}^D_r = \emptyset$ 
for every $1 \leq r \leq c$, which implies that the second multiplicative 
factor for $s_{CCA}$ is $O(1)$. Hence $s_{CCA} = O(1)$ is an order of 
magnitude smaller than $|E_\sigma| = O(p)$. 

\medskip

\noindent
{\it Example 2 (cycle) contd.} If $\sigma$ is the RCM fill reducing 
ordering (see Figure \ref{cycle}), it then follows that the maximum 
number of non-zero entries in any row or column of $\bar{L}$ is bounded 
by 4, hence $a^D_n = O(1)$, and $\min(a^D_n, |E^D_\sigma|) = O(1) << p = 
|E_\sigma|$. However, the second multiplicative factor for $s_{CCA}$ 
turns out to be exponential in $p$. In this case, $c = p-3$, and 
straightforward calculations show that $\max_{1 \leq r \leq c} g^D (r) = 
g^D (c) = O \left( (3 \delta^{-1})^{p-4} \right)$ (see Assumption A1 
above). Hence, overall $|E_\sigma|$ is of a much smaller order than $s_{CCA}$. {\color{black} We show however that through more targeted arguments specific to the $p$-cycle setting, and an additional constraint on $\bar{L}$, a significantly better convergence rate than $\sqrt{s_{CCA} \log p/n}$ can be obtained for the CCA estimator. In particular, for proving such a result, Assumption A3 above needs to be replaced by the following assumption. 
\begin{itemize}
\item ($A3^\prime$) The quantity $\max_{3 \leq j \leq p-1} \frac{|\bar{L}_{j,j-2}|}{\bar{L}_{jj}}$ is uniformly (in $n$) bounded away from 1, and $\frac{|E_\sigma| \log p}{n} \rightarrow 0$ as $n \rightarrow \infty$. 
\end{itemize}

\noindent
The first part of Assumption $A3^\prime$ essentially constrains the non fill-in entry in most rows of $L_0$ to be smaller in magnitude than the corresponding diagonal entry. The following lemma now establishes a Frobenius norm convergence rate for the CCA estimator that matches the Frobenius norm convergence rate for the grahical Gaussian MLE. The proof of this lemma is provided in the Supplemental material at the end of the paper. 
\begin{lem} (Improved CCA bounds for $p$-cycle)
Consider the setting where $G_\sigma = (V, E_\sigma)$ is a $p$-cycle 
equipped with the RCM based fill reducing ordering illustrated in 
Figure \ref{cycle}. Then, under Assumptions A1, A2 and $A3^\prime$, there exist 
constants $\tilde{K}$ and $\tilde{K}'$ such that the estimators $\hat{L}$ and 
$\hat{\Omega}$ satisfy 
		$$
		\bar{P} \left( \left\{\norm{\widehat{L} - \bar{L}}_F \leq \tilde{K} \sqrt{\frac{(|E_\sigma|+1) \log p}{n}} \right\} \cap \left\{ \norm{\widehat{\Omega} - \bar{\Omega}}_F 
        \leq \tilde{K}' \sqrt{(\frac{|E_\sigma|+1) \log p}{n}} \right\} \right) \rightarrow 1 
		$$

\noindent
as $n \to \infty$. 
\end{lem}}

\medskip

\noindent
{\it Example 3 (almost complete graph) contd.} In this case both 
$\min(a^D_n, |E^D_\sigma|)$ and $|E_\sigma|$ are of order $p^2$, and 
since there is only fill-in entry ($c=1$), it follows that the second 
multiplicative factor for $s_{CCA}$ is $O(1)$. Hence, in this case both 
$s_{CCA}$ and $|E_\sigma|$ are of order $p^2$. 

\medskip

\noindent
{\it Example 4 (Bipartite graph) contd.} Consider a bipartite graph 
with sets $A$ and $B$ forming a partition of the vertex set $V$, such 
that $|A|=m$ and $|B|=p-m$ with $m = O(1)$. Consider the ordering 
$\sigma$ as defined in Section \ref{sec:complexity} for this setting. 
In this case both $\min(a^D_n, |E^D_\sigma|)$ and $|E_\sigma|$ are of 
order $p$. Again, since $c = {{m}\choose{2}} = O(1)$, it follows that the 
second multiplicative factor for $s_{CCA}$ is $O(1)$. Hence, in this case 
both $s_{CCA}$ and $|E_\sigma|$ are of order $p$. 

\medskip

\noindent
{\it Example 5 (Multipartite graph) contd.} In this case, both 
$\min(a^D_n, |E^D_\sigma|)$ and $|E_\sigma|$ are of order $p^2$. Since 
fill-in edges are only allowed within each $3$-element partition sets 
$A_i$, it can be shown that $\max_{1 \leq r \leq c} g^D (r)$ and 
$\tilde{a}^D_n$ are $O(1)$. To conclude, $s_{CCA}$ and $|E_\sigma|$ 
are both order $p^2$.}

\medskip

\noindent
In summary, the examples above demonstrate that excpet in settings when the underlying graphs 
are highly sparse (such as cycles), the proposed CCA approach has demonstrably better 
statistical performance than the MLE (in terms of best available operator norm convergence rate)
for moderately sparse and dense graphs. This combined with its superior computational complexity 
suggests that in modern high-dimensional settings, the CCA approach should be the preferred 
method. The only exceptions are cases when the underlying graph is a cycle or a graph close to a cycle.

\section{Numerical Illustrations and real data applications} \label{sec:experiments}

\subsection{Simulation experiment: {\color{black} evaluation of {CCA} when $G$ is given}} \label{sec:simulated:data}

\noindent
In this section, we evaluate the high-dimensional empirical performance of the CCA algorithm in a simulation setting where the underlying graph $G$ is assumed to be given/known. Against the backdrop of the results on computational complexity in Section \ref{sec:complexity} (Lemma \ref{lem_CF}, Lemma \ref{lem_CG} and Table \ref{computational:complexity}) we compare the performance of the {CCA} algorithm and the {G-IPF}  algorithm. For each value of $p \in \{500,1000,1500, 2000,5000\}$, the following procedure is repeated: 
The true $\Omega$ has a random sparsity pattern of approximately $2p$ edges. To generate $\Omega = L^tL$ for a given 
$p$, we initally generate $L$, a lower triangular matrix, in the following manner: Randomly select some entires of L to be 
zero (approximately less than $2p$). Amongst these we randomly set half of them to be positive and the other half to be negative. 
In addition, the off-diagonal entries of $L$ are drawn from the uniform distribution over $[0.3,0.7]$. Finally, we choose the 
diagonal entries from the uniform distribution over $[2,5]$. The sample size takes values as follows: $n = \{0.5p, 0.75p, p, 2p\}$. We generate 50 datasets for each of these settings from both the 
multivariate Gaussian and the heavy-tailed multivariate-$T$ distribution with 3 degrees of 
freedom. 

For the Gaussian case, both the {CCA} algorithm and the {G-IPF}  algorithm are then used to obtain estimates of 
$\Omega$. We also present results from a hybrid approach, which uses the {CCA} estimate as the initial estimate 
in G-IPF as opposed to the diagonal matrix with diagonal entries $\{1/S_{ii}\}_{i=1}^p$. The results presented below are the 
average relative Frobenius norm estimation errors ($\frac{\norm{\hat{\Omega}-\Omega}}{\norm{\Omega}}$) over the 50 
datasets generated for each sample size and underlying distribution. The results of the simulations for the multivariate Gaussian case are provided in Table \ref{Gaussian}. CCA produces significantly better wall-times compared to G-IPF without compromising on accuracy. 

For the multivariate-$T$ datasets, we compare the performance of the {CCA} algorithm to {G-IPF}, and the {\it tlasso-IPF} 
algorithm from \cite[Section 3.2]{Feingold:Drton:2011}. The results are provided in Table \ref{tdist}. The results show 
that the scalability of the {CCA} algorithm is even more pronounced compared to {G-IPF}  in the heavy-tailed 
$T$-distribution case. Note that the {\it tlasso-IPF} uses the EM algorithm to obtain maximum likelihood estimates of 
concentration matrices (with a given sparsity pattern) assuming that the underlying data is generated from a 
multivariate-$t$ distribution. It is a double iterative algorithm, as the M-step in each iteration of the EM algorithm 
involves the iterative IPF algorithm. The maximum number of iterations for the 
EM algorithm was set to $100$, and for the IPF algorithm was set to $1000$. We report an ``N/A" for estimation error values of the iterations for a given setting which does not finish in $4$ days. For $p = 2000$ and $5000$, this is the case for the {\it t-lasso} algorithm. For lower values of $p$, the time taken by {\it t-lasso} is anywhere between $500$ to $2000$ times more as compared to the {CCA} algorithm. The estimation error for the resulting estimates is still very high, which suggests that more iterations are needed for convergence of the EM algorithm. To summarize, the results in Tables \ref{Gaussian} and 
\ref{tdist} demonstrate how state-of-the-art iterative algorithms can be prohibitively 
expensive in high-dimensional settings, and provide compelling reasons to use the proposed 
non-iterative {CCA} algorithm for fast and accurate estimation in modern high-dimensional applications. 
\begin{table}
	\centering
	{\footnotesize \begin{tabular}{|r|r|rr|rr|rr|c|}
		\hline
		& & \multicolumn{2}{|c|}{CCA} & \multicolumn{2}{|c|}{G-IPF} & \multicolumn{2}{|c|}{CCA-GIPF} & { Improvement Factor}\\
		p & n & Time & Norm & Time & Norm &Time & Norm & { (G-IPF/CCA Time)}\\ 
		\hline
		500 & 250 & {5.2} & {0.1005} & 28.6 & 0.1014 & 10.1 & 0.1009 & {\bf 5.5}\\ 
		500 & 375 & { 5.3} & 0.0817 & 28.7 & 0.0822 & 10.1 & 0.0819 & {\bf 5.4}\\ 
		500 & 500 & { 5.3} & 0.0700 & 27.9 & 0.0703 & 9.4 & 0.0701 & {\bf 5.3}\\ 
		500 & 1000 & { 5.3} & 0.0492 & 24.2 & 0.0493 & 8.6 & 0.0493 & {\bf 4.6}\\ 
		\hline
		1000 & 500 & { 34.8} & 0.0762 & 536.3 & 0.0765 & 340.2 & 0.0762 & {\bf 15.4}\\ 
		1000 & 750 & { 34.8} & 0.0626 & 516.5 & 0.0627 & 332.8 & 0.0625 & {\bf 14.8}\\ 
		1000 & 1000 & { 34.5} & 0.0536 & 513.2 & 0.0537 & 333.5 & 0.0536 & {\bf 14.9}\\ 
		1000 & 2000 & { 35.3} & 0.0376 & 495.1 & 0.0377 & 306.8 & 0.0376 & {\bf 14.0}\\ 
		\hline
		1500 & 750 & { 112.7} & 0.0648 & 2661.3 & 0.0643 & 1885.0 & 0.0641 & {\bf 23.6}\\ 
		1500 & 1125 & { 112.9} & 0.0529 & 2622.1 & 0.0525 & 1841.2 & 0.0524 & {\bf 23.1}\\ 
		1500 & 1500 & { 113.0} & 0.0457 & 2580.0 & 0.0453 & 1809.0 & 0.0452 & {\bf 22.8}\\ 
		1500 & 3000 & { 116.0} & 0.0322 & 2471.5 & 0.0318 & 1737.1 & 0.0319 & {\bf 21.3}\\ 
		\hline
		2000 & 1000 & { 289.8} & 0.0555 & 8933.6 & 0.0551 & 5179.3 & 0.0549 & {\bf 30.8}\\ 
		2000 & 1500 & { 298.9} & 0.0454 & 8844.7 & 0.0450 & 5112.5 & 0.0449 & {\bf 29.5}\\ 
		2000 & 2000 & { 299.5} & 0.0391 & 8740.4 & 0.0387 & 5016.7 & 0.0387 & {\bf 29.2}\\
		2000 & 4000 & { 309.3} & 0.0276 & 8590.3 & 0.0273 & 4870.1 & 0.0273 & {\bf 27.8}\\
		\hline
		5000 & 2500 & { 4911.3} & 0.0320 & 120252.4 & 0.0321 & 33776.3 & 0.0320 & {\bf 24.5}\\
		5000 & 3750 & { 4945.8} & 0.0261 & 113356.6 & 0.0261 & 28047.8 & 0.0261 & {\bf 22.9}\\
		5000 & 5000 & { 5034.5} & 0.0226 & 114867.7 & 0.0226 & 29037.7 & 0.0226 & {\bf 22.8}\\
		5000 & 10000 & { 5156.2} & 0.0160 & 114133.3 & 0.0160 & 29248.7 & 0.0160 & {\bf 22.1}\\
		\hline
	\end{tabular}}
	\caption{Running time and estimation error (in relative Frobenius norm) for CCA, {G-IPF}, 
		and CCA-GIPF (use {G-IPF}  with the {CCA} estimator as the initial value) with normal data. {\color{black} All 
		methods have very similar estimation errors, but {CCA} is significantly faster. The ratio of running times of {G-IPF}  to {CCA} is provided in bold.}}
	\label{Gaussian}
\end{table}

\begin{table}
	\centering
	{\footnotesize \begin{tabular}{|r|r|rr|rr|rr|c|}
		\hline
		&  & \multicolumn{2}{|c|}{CCA} & \multicolumn{2}{|c|}{G-IPF} & \multicolumn{2}{|c|}{{\it tlasso-IPF}} & { Improvement Factor}\\
		p & n & Time & Norm & Time & Norm &Time & Norm & { ({\it {G-IPF} }/CCA Time)}\\
		\hline
		500 & 250 & {4.0} & 0.369 & 54.41 & 0.372 & 1933.06 & 6.862 & {\bf 13.6}\\
		500 & 375 & { 4.0} & 0.280 & 36.26 & 0.282 & 1792.392 & 6.817 & {\bf 9.1}\\
		500 & 500 & { 3.9} & 0.238 & 37.52 & 0.239 & 1772.345 & 6.779 & {\bf 9.6}\\
		500 & 1000 & { 3.9} & 0.232 & 37.27 & 0.232 & 1891.426 & 6.753 & {\bf 9.6}\\
		\hline
		1000 & 500 & { 27.0} & 0.322 & 1548.9 & 0.323 & 36508.994 & 8.878 & {\bf 57.4}\\
		1000 & 750 & { 26.9} & 0.217 & 719.9 & 0.218 & 35761.71 & 8.846 & {\bf 26.8}\\
		1000 & 1000 & { 27.9} & 0.208 & 650.1 & 0.208 & 35364.331 & 8.838 & {\bf 23.3}\\
		1000 & 2000 & { 28.8} & 0.170 & 677.4 & 0.170 & 34951.69 & 8.802 & {\bf 23.5}\\
		\hline
		1500 & 750 & { 95.193} & 0.258 & 6807.133 & 0.258 & 185566.52 & 10.231 & {\bf 71.5}\\
		1500 & 1125 & { 95.517} & 0.190 & 3559.467 & 0.189 & 185772.6 & 10.195 & {\bf 37.7}\\
		1500 & 1500 & { 92.459} & 0.196 & 3846.347 & 0.195 & 180279.2 & 10.182 & {\bf 41.6}\\
		1500 & 3000 & { 94.754} & 0.153 & 3392.462 & 0.152 & 174894.6 & 10.166 & {\bf 35.8}\\
		\hline
		2000 & 1000 & { 234.100} & 0.228 & 12738.264 & 0.219 & $> 4$ days & NA & {\bf 54.4}\\
		2000 & 1500 & { 263.570} & 0.200 & 17389.751 & 0.200 & $> 4$ days & NA & {\bf 66.0}\\
		2000 & 2000 & { 308.789} & 0.172 & 10757.530 & 0.172 & $> 4$ days & NA & {\bf 34.8}\\
		2000 & 4000 & { 356.389} & 0.143 & 11920.13 & 0.142 & $> 4$ days & NA & {\bf 33.4}\\
		\hline
		5000 & 2500 & { 5019.258} & 0.165 & 163465.7 & 0.139 & $> 4$ days & NA & {\bf 32.6}\\
		5000 & 3750 & { 4817.014} & 0.118 & 155700.3 & 0.118 & $> 4$ days & NA & {\bf 32.3}\\
		5000 & 5000 & 5051.542 & 0.140 & 159974.9 & 0.131 & $> 4$ days & NA & {\bf 31.7}\\
		5000 & 10000 & 4594.165 & 0.097 & 139444.2 & 0.097 & $> 4$ days & NA & {\bf 30.4}\\
		\hline
	\end{tabular}}
	\caption{Running time and estimation error (in relative Frobenius norm) for CCA, {G-IPF} , 
		and {\it tlasso-IPF} with $t$ data. {\color{black} {G-IPF}  and {CCA} again have very similar estimation 
		errors, but {CCA} is significantly faster. The ratio of running times of {G-IPF}  to {CCA} is provided in bold, as 
		{\it tlasso-IPF} values are not available for some of the settings.}}
	\label{tdist}
\end{table}

{\color{black}
\subsection{Simulation experiment: evaluation of {CCA} when $G$ is not given}\label{sec:simulated:second}

In practice, the graph $G$ is not given/known. In this setting, a graph selection step 
is required prior to CCA, and we use the {\it FST} method \citep{zhang2021quadratic} for this purpose. {\it FST} is a non-iterative thresholding method (extending ideas in \cite{rothman2009generalized}) that is 
scalable to ultra high-dimensional settings, and is also theoretically supported by high-dimensional statistical 
consistency results (see Section 5 of \cite{zhang2021quadratic} for details). However, the sparse estimate of 
$\Omega$ provided by {\it FST} is not guaranteed to be positive definite. Setting $G$ as the sparsity pattern in 
$\Omega$ estimated by FST, and then applying the proposed {CCA} method in Section \ref{sec:congraphmd} yields 
a positive definite estimate of $\Omega$ with exactly the same sparsity pattern as the {\it FST} estimate of 
$\Omega$. We refer to this approach as {\it FST+CCA}.  The {\it FST} based estimate of $G$ can have several 
connected components, in which case the {CCA} method is separately applied on each component. Doing so enables us to use 
techniques such as sparse matrix representation and parallel computation to gain efficiency. 

We compare the {\it FST+CCA} method to two leading methods: Glasso and BigQUIC. These two are state-of-the-art penalized likelihood methods for simultaneous sparsity selection and positive definite estimation, and BigQUIC in 
particular is scalable to the ultra high-dimensional $p=10^5$ setting considered in this section. Implementations 
of both these methods are available in {\it R}. Simulations comparing the {\it FST+CCA} method with 
Glasso consider the setting $p = \{1000, 2000\}$ with $n = 0.5p$, and with BigQUIC use $p = \{10^4, 10^5 \}$, with $n = 1000$. While Glasso is able to comfortably handle settings with a couple of thousand variables, it can be really expensive to 
allocate enough memory for Glasso to scale to $p>10^4$. Both the dense matrix 
representations used in the internal steps and the fact that Glasso cannot be straightforwardly parallelized using multiple cores hinder its space and time scalability to ultra-high dimensions. BigQUIC, on the other hand, is able to scale to $p>>10^4$ by employing several innovations which include performing block-wise updates and the use of a fast approximation for the Hessian (see detailed discussion in \cite{hsieh2013big}). As the algorithm is iterative in nature, a `bad' partition of the variables during blocking can hinder its performance. Here we perform the comparative study of the {\it FST+CCA} method in a single core situation with Glasso when $p$ is moderately large, and in a multi-core situation with BigQuic with ultra large $p$.

For each choice of $p$, a true $\Omega$ is obtained by the random block model (RBM) procedure and also by a random banded line graph (RBL) design procedure, which randomly generates the sub-diagonal entries in the precision matrix that corresponds to a line graph. Both these are described in detail in 
\cite[Section5]{zhang2021quadratic}. 
Once this is done, we simulate multivariate Gaussian data with the given true $\Omega$ as the precision matrix. FST is then applied to this data to determine the graph structure and {CCA} is applied in parallel to each connected 
component resulting from {\it FST} to obtain the final positive definite estimate of $\Omega$. Tuning parameters for FST, Glasso and BigQUIC are determined using five fold cross validation. A total of $50$ datasets 
are generated for each case and the average errors in relative Frobenius norm and average running time are 
provided in Table \ref{tab:glasso:summary}, \ref{tab:small_dimension:quic}, and \ref{tab:quic:summary}. 

\begin{table}
\centering
	\begin{tabular}{|c|r|r|rr|rr|c|}
		\hline 
		\multirow{4}{*}{RBM} &  &  & \multicolumn{2}{c|}{FST+CCA} & \multicolumn{2}{c|}{Glasso} &  
		Improvement Factor\\
		& p & n & Time & Norm & Time & Norm & $\frac{\text{Glasso}}{\text{FST+CCA}}$\tabularnewline
		\cline{2-8} \cline{3-8} \cline{4-8} \cline{5-8} \cline{6-8} \cline{7-8} \cline{8-8} 
		& 1000 & 500 & 2.90 secs & 0.206 & 1.44 mins & 0.225 & {\bf 29.79} times\\
		\cline{2-8} \cline{3-8} \cline{4-8} \cline{5-8} \cline{6-8} \cline{7-8} \cline{8-8} 
		& 2000 & 1000 & 21.01 secs & 0.124 & 14.74 mins & 0.177 & {\bf 42.09} times\\
		\hline 
	\end{tabular}
	
	\begin{tabular}{|c|r|r|rr|rr|c|}
		\hline 
		\multirow{4}{*}{RBL} &  &  & \multicolumn{2}{c|}{FST+CCA} & \multicolumn{2}{c|}{Glasso} &  
		Improvement Factor\\
		& p & n & Time & Norm & Time & Norm & $\frac{\text{Glasso}}{\text{FST+CCA}}$\\
		\cline{2-8} \cline{3-8} \cline{4-8} \cline{5-8} \cline{6-8} \cline{7-8} \cline{8-8} 
		& 1000 & 500 & 1.25 secs & 0.096 & 2.27 mins & 0.241 & {\bf 108.96} times\\
		\cline{2-8} \cline{3-8} \cline{4-8} \cline{5-8} \cline{6-8} \cline{7-8} \cline{8-8} 
		& 2000 & 1000 & 31.18 secs & 0.059 & 18.19 mins & 0.201 & {\bf 35.00} times\\
		\hline 
	\end{tabular}
	
	\caption{Running time and estimation error (in relative Frobenius norm) for {\it FST+CCA} and Glasso, in both RBM and RBL designs, using 1 core. {\color{black} {\it FST+CCA} is more accurate and significantly faster than Glasso for both designs. The ratio of running times of Glasso to {\it FST+CCA} is provided in bold.}}
	\label{tab:glasso:summary}
\end{table}

In Table \ref{tab:glasso:summary}, we provide results for comparison of {\it FST+CCA} with Glasso in the $p \in 
\{1000,2000\}$ setting. The Glasso implementation available in {\it R} cannot use parallel processing via 
multiple cores. While {\it FST+CCA} has this parallelizable property, for comparison purposes we employ only a single core for all the experiments in 
Table \ref{tab:glasso:summary}. Even without the benefit of parallel processing, {\it FST+CCA} is at least 30 to 100 times faster than Glasso, and performs competitively or better than Glasso in terms of accuracy in all settings. 

\begin{table}
    \centering
    \begin{tabular}{|c|r|r|rr|rr|c|}
        \hline 
        \multirow{4}{*}{RBM} &  &  & \multicolumn{2}{c|}{FST+CCA} & \multicolumn{2}{c|}{BigQuic} &  
        Time Improvement Factor\\
        & p & n & Time & Norm & Time & Norm & $\frac{\text{BigQuic}}{\text{FST+CCA}} $\tabularnewline
        \cline{2-8} \cline{3-8} \cline{4-8} \cline{5-8} \cline{6-8} \cline{7-8} \cline{8-8} 
        & 1000 & 500 & 0.77 secs & 0.204 & 1.06 secs & 0.286 & {\bf 1.38} times\\
        \cline{2-8} \cline{3-8} \cline{4-8} \cline{5-8} \cline{6-8} \cline{7-8} \cline{8-8} 
        & 2000 & 1000 & 2.47 secs & 0.133 & 6.76 secs & 0.284 & {\bf 2.73} times\\
        \hline 
    \end{tabular}
        \begin{tabular}{|c|r|r|rr|rr|c|}
        \hline 
        \multirow{4}{*}{RBL} &  &  & \multicolumn{2}{c|}{FST+CCA} & \multicolumn{2}{c|}{BigQuic} &  
        Time Improvement Factor\\
        & p & n & Time & Norm & Time & Norm & $\frac{\text{BigQuic}}{\text{FST+CCA}} $\tabularnewline
        \cline{2-8} \cline{3-8} \cline{4-8} \cline{5-8} \cline{6-8} \cline{7-8} \cline{8-8} 
        & 1000 & 500 &0.43 secs &  {0.151} & 1.44 secs & {0.303} & {\bf 3.35} times\\
        \cline{2-8} \cline{3-8} \cline{4-8} \cline{5-8} \cline{6-8} \cline{7-8} \cline{8-8} 
        & 2000 & 1000 & 2.67 secs & {0.074}  & 9.46 secs & {0.312} & {\bf 3.54} times\\
        \hline 
    \end{tabular}
    \caption{Running time and estimation error (in relative Frobenius norm) for {\it FST+CCA} and BigQUIC, with $p = 1000, 2000$, in both RBM and RBL designs, using 16 cores. {\color{black} {\it FST+CCA} is more accurate and faster than BigQUIC for both designs. The ratio of running times of BigQUIC to {\it FST+CCA} is provided in bold.}}
    \label{tab:small_dimension:quic}
\end{table}

Table \ref{tab:small_dimension:quic} provides results for comparing {\it FST+CCA} with BigQuic in the $p \in 
\{1000,2000\}$ setting, using 16 cores. It shows that {\it FST+CCA} yields lower wall times  and delivers an even better advantage in accuracy. In Table \ref{tab:quic:summary}, we provide results for comparison of {\it FST+CCA} with BigQUIC in the $p \in 
\{10^4,10^5\}$ setting. As mentioned previously, the Glasso implementation in R can break down in these 
settings due to the use of dense matrix representations. Both {\it FST+CCA} and BigQUIC however have the ability to use 
multiple cores for parallel processing, and each method is provided with 16 cores for the experiments in 
Table \ref{tab:quic:summary}. In all settings, {\it FST+CCA} is at least 5 to 12 times faster than 
BigQUIC, and can also substantially improve accuracy compared to BigQUIC. To summarize, the simulations in 
this section illustrate that the proposed {\it FST+CCA} can be significantly faster than state of the art ultra high-dimensional iterative approaches, while providing better estimation accuracy. The results above also underscore how CCA can be readily coupled with high-dimensional sparsity selection methods. 
\begin{table}
	\begin{tabular}{|c|r|r|rr|rr|c|}
		\hline 
		\multirow{4}{*}{RBM} &  &  & \multicolumn{2}{c|}{FST+CCA} & \multicolumn{2}{c|}{BigQUIC} & 
		Time Improvement Factor\\
		& p & n & Time & Norm & Time & Norm & (BigQUIC/FST+CCA Time)\\
		\cline{2-8} \cline{3-8} \cline{4-8} \cline{5-8} \cline{6-8} \cline{7-8} \cline{8-8} 
		& $10^4$ & 1000 & 13.39 secs  & 0.062 & 105.13 secs & 0.288  & {\bf 7.85} times \\
		\cline{2-8} \cline{3-8} \cline{4-8} \cline{5-8} \cline{6-8} \cline{7-8} \cline{8-8} 
		& $10^5$& 1000 & 29.77 mins & 0.144 & 183.61 mins & 0.150 & {\bf 6.16} times\\
		\hline 
	\end{tabular}
	
	\begin{tabular}{|c|r|r|rr|rr|c|}
		\hline 
		\multirow{4}{*}{RBL} &  &  & \multicolumn{2}{c|}{FST+CCA} & \multicolumn{2}{c|}{BigQUIC} & 
		Time Improvement Factor\\
		& p & n & Time & Norm & Time & Norm & (BigQUIC/FST+CCA Time)\\
		\cline{2-8} \cline{3-8} \cline{4-8} \cline{5-8} \cline{6-8} \cline{7-8} \cline{8-8} 
		& $10^4$ & 1000 & 16.65 secs  & 0.041 & 196.45 secs & 0.301 & {\bf 11.80} times\\
		\cline{2-8} \cline{3-8} \cline{4-8} \cline{5-8} \cline{6-8} \cline{7-8} \cline{8-8} 
		& $10^5$& 1000 & 31.01 mins & 0.087 & 151.70 mins & 0.319 & {\bf 4.76} times\\
		\hline 
	\end{tabular}

	\caption{Running time and estimation error (in relative Frobenius norm) for {\it FST+CCA} and BigQUIC, with $p = 10^4, 10^5$, in both RBM and RBL designs, using 16 cores. {\color{black} {\it FST+CCA} is more accurate and faster than BigQUIC for both designs. The ratio of running times of BigQUIC to {\it FST+CCA} is provided in bold.}}
	\label{tab:quic:summary}
\end{table}}

\subsection{Application to minimum variance portfolio rebalancing} 

\subsubsection{S\&P 500 data with 398 stocks when $G$ is given (CCA vs. G-IPF)} \label{sec:finance:data}

\noindent
The minimum variance portfolio selection problem is defined as follows. Given $p$ risky assets, let $r_{ik}$ denote the return 
of asset $i$ over time period $k$ (for $i = 1,2, \cdots, p$); which in turn is defined as the change in its price over time period $k$, divided by the price 
at the beginning of the period. Let $r^T_k = (r_{1k}, r_{2k}, \cdots, r_{pk})$ denote the vector of returns over time period $k$. 
Let $w^T_k = (w_{1k}, w_{2k}, . . . , w_{pk})$ denote the vector of portfolio weights, i.e., $w_{ik}$ denotes the weight of asset 
$i$ in the portfolio for the $k$-th time period. A long position or a short position for asset $i$ during period $k$ is given by the 
sign of $w_{ik}$, i.e., $w_{ik} > 0$ for long, and $w_{ik} < 0$ for short positions respectively. The budget constraint can be 
written as $\mathbf{1}^Tw_k = 1$, where $\mathbf{1}$ denotes the vector of all ones. Let $\Sigma$ denote the covariance 
matrix of the vector of returns. Note that the risk of a given portfolio, as measured by the standard deviation of its return, is 
simply $(w^T_k \Sigma_k w_k)^{1/2}$. The minimum variance portfolio selection problem for investment period $k$ can now 
be formally defined as follows:
$\min_{w_k} w^T_k \Sigma_k w_k \text{ subject to } \mathbf{1}^Tw_k = 1$. As the minimum variance portfolio problem is a simple quadratic program, it has an analytic solution $w^*_k = (\mathbf{1}^T \Sigma_k^{-1}  \mathbf{1})^{-1} 
\Sigma_k^{-1} \mathbf{1}$. The most basic approach to the portfolio selection problem often makes the unrealistic 
assumption that returns are stationary in time. A standard approach to dealing with the non-stationarity in such financial time 
series is to use a periodic rebalancing strategy. In particular, at the beginning of each investment period $k = 1, 2, . . . , K$, 
portfolio weights $w^T_k = (w_{1k}, w_{2k}, \dots , w_{pk})$ are computed from the previous $N_{est}$ days of observed returns ($N_{est}$ is called the estimation horizon). These portfolio weights are then held constant for the duration of each 
investment period. The process is repeated at the start of the next investment period and is often referred to as ``rebalancing". 

We now consider the problem of investing in the stocks that feature in the S\&P 500 index. The S\&P 500 is a composite 
blue chip index consisting of 500 stocks. A number of stocks were removed in our analysis due to limited data 
span of the corresponding companies as constituents of the S\&P 500 index - we ultimately used 398 stocks. For illustration purposes, rebalancing time points were chosen to be every 64 weeks starting from 2000/01/01 to 2016/12/31 (approximately 17 
years). Start and end dates of each period are selected to be calendar weeks, and need not coincide with a trading day. The 
total number of investment periods is 14, and the number of trading days in each investment period is approximately 320 
days. We shall compare the performance of two approaches, ones which use {G-IPF}  and {CCA} for estimating the covariance matrix at each rebalancing point. We consider various 
choices of $N_{est}$, in particular, $N_{est} \in {375,450,525}$ in our analysis. Note that once a choice for $N_{est}$ is 
made, it is kept constant until the next investment period.

Note that for {G-IPF}  and {CCA} a graph for the zeros in the concentration matrix is required. When $N_{est}$ 
is less than $398$, we threshold the generalized inverse for the sample covariance, and when $N_{est}$ is larger than 398, we 
threshold the inverse till we achieve a sparsity of approximately 95\%. The {\it Normalized wealth growth} is defined as the 
accumulated wealth derived from the portfolio over the trading period when the initial budget is normalized to one. This 
measure is also designed to take into account both transaction costs and borrowing costs. Figure \ref{fig:norm375} shows that in terms of 
Normalized Wealth Growth, the {G-IPF} and {CCA} based approaches perform very similarly and both perform better than the passive S\&P500 index approach (no rebalancing) when $N_{est} = 375$. Very similar results are obtained for $N_{est} = 450$ and $525$ (not illustrated here due to space considerations). Overall, the results 
demonstrate that the {CCA} and {G-IPF}  based strategies are very competitive with each other in terms of financial 
performance. However, as demonstrated in Table \ref{tab:time}, the {CCA} algorithm is much faster in terms of computational 
time compared to the {G-IPF}  approach. Since the inverse covariance matrix has to be estimated at the start of each 
investment horizon, the gain in computational speed can add up very quickly. The above results reinforce our 
earlier observations in Section \ref{sec:simulated:data} that the {CCA} algorithm provides a significantly more 
scalable alternative to the {G-IPF}  algorithm, while providing similar statistical performance. 

\begin{figure}
\begin{floatrow}
\ffigbox{%
  \includegraphics[width=3in,height=3in]{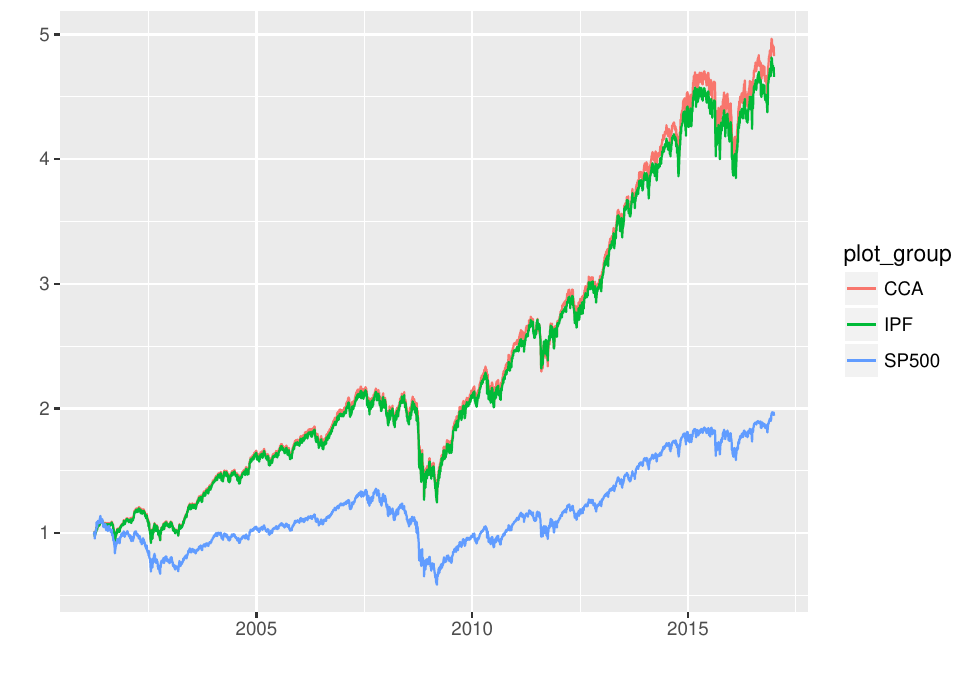}
}{%
  
  \caption{Normalized wealth growth after adjusting for transaction costs (0.5\% of principal)
		and borrowing costs (interest rate of 8\% APR) with $N_{est} = 375$ for 
  S\&P 500 data.}
  \label{fig:norm375}
}
\capbtabbox{%
  \begin{tabular}{r|rrr}
				\hline
				& & $N_{est}$ &\\
				\hline
				Method & 375 & 450 & 525 \\ 
				\hline
				CCA  & 1.92 & 1.74 & 1.67 \\ 
				G-IPF & 3.77 & 4.77 & 5.31 \\ 
				\hline
			\end{tabular}
}{%
  
		\caption{Average running time (in seconds) for estimating the covariance matrix of returns with {CCA} and {G-IPF}  for various 
			values of $N_{est}$ for S\&P 500 data}
   \label{tab:time}
}
\end{floatrow}
\end{figure}

\subsubsection{Kaggle data with 3630 stocks/ETFs when graph $G$ is unknown} \label{sec:finance:fstcca}

\noindent
In this section, we consider the minimum variance portfolio problem when investing in a much larger number of stocks. The dataset is obtained from an open source publication on Kaggle, provided by \cite{HugeStock}. Specifically 3630 stocks and ETFs which have a complete record from 2014-06-09 to 2017-06-07 (featuring 756 trading days) are selected. Daily returns are calculated. Similar to the previous subsection, we use a rebalancing scheme in this application. A significant difference to address here is that we have fewer number of observations but a much larger number of variables in this dataset. Thus, it is reasonable to choose a relative smaller rebalancing time and larger $N_{est}$. We consider the choice of $N_{est} = 500$, and rebalancing time of $30$ days. This will give us a portfolio to be used from days 501 to 756, which correspond to the period 2016-06-02 to  2017-06-07. More specifically, at beginning, the first 500 daily returns are used to construct a portfolio which will be used in days 501 to 530. Then for days 531 to 560, data from days 31 to 530 will be used for an updated portfolio, and so forth. This scheme gives us a total of 9 time-points for portfolio rebalancing. Here we compare the performance of three approaches, which respectively use {\it FST+CCA}, Glasso and BigQuic for estimating the precision matrix at each rebalancing point.  

For the selection of tuning hyper-parameters, we used the portfolio loss as the loss function and a validation set of size 10. Specifically, for 500 daily return observations, 490 observations were used to construct the portfolio for different tuning parameter values and the last 10 days of portfolio returns were used to determine the best tuning parameter.

The average running time for all the three approaches are recorded in Table \ref{tab:large_stock_data:time}. From Table \ref{tab:large_stock_data:time}, {\it FST+CCA} takes about $1/5$ of the time required by BigQuic, and less than $1/200$ of the time used by Glasso, demonstrating the efficiency of the {\it FST+CCA} approach.

\begin{table}[ht]
\begin{center}
\begin{tabular}{c|c|c|c}
\hline
Approaches             & {\it FST+CCA} & BigQuic & Glasso  \\ \hline
Running Time (Seconds) & 6.883   & 34.531 & 1497.781 \\ \hline
\end{tabular}
\end{center}
\label{tab:large_stock_data:time}
\caption{Average running time (in seconds) for estimating the covariance matrix of returns with FST+CCA, Glasso and BigQuic, using data on 3630 stocks and ETFs from \cite{HugeStock}}
\end{table}

The normalized wealth growth derived from the portfolios is illustrated in Figure \ref{fig:large_stock_data:cumwealth}, and shows that the {\it FST+CCA} approach has better overall wealth growth. The daily returns of these portfolios are presented in Figure \ref{fig:large_stock_data:return}, demonstrating that the {\it FST+CCA} approach is able to provide a portfolio which is less volatile compared to the other approaches. Overall, {\it FST+CCA} performs better than the competing methods in terms of both computational efficiency and wealth growth of the associated portfolio. 

\begin{figure}
	\begin{center}
		\includegraphics[width=6.5in,height=3.5in]{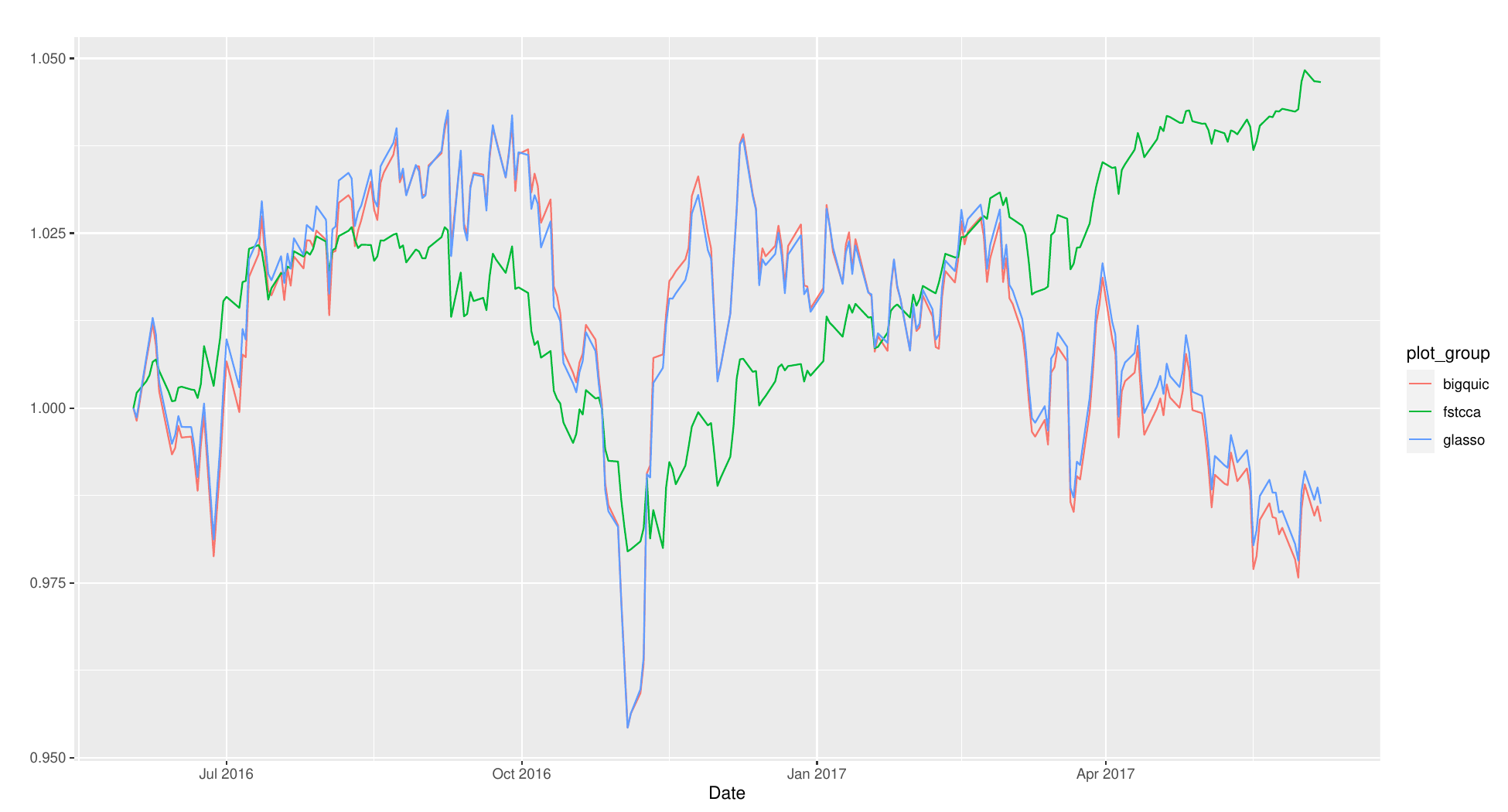}
	\end{center}
	\caption{Normalized wealth growth after adjusting for transaction and borrowing costs with $N_{est} = 500$ for stock/ETF data from \cite{HugeStock}.}
 \label{fig:large_stock_data:cumwealth}
\end{figure}

\begin{figure}
	\begin{center}
		\includegraphics[width=6.5in,height=3.5in]{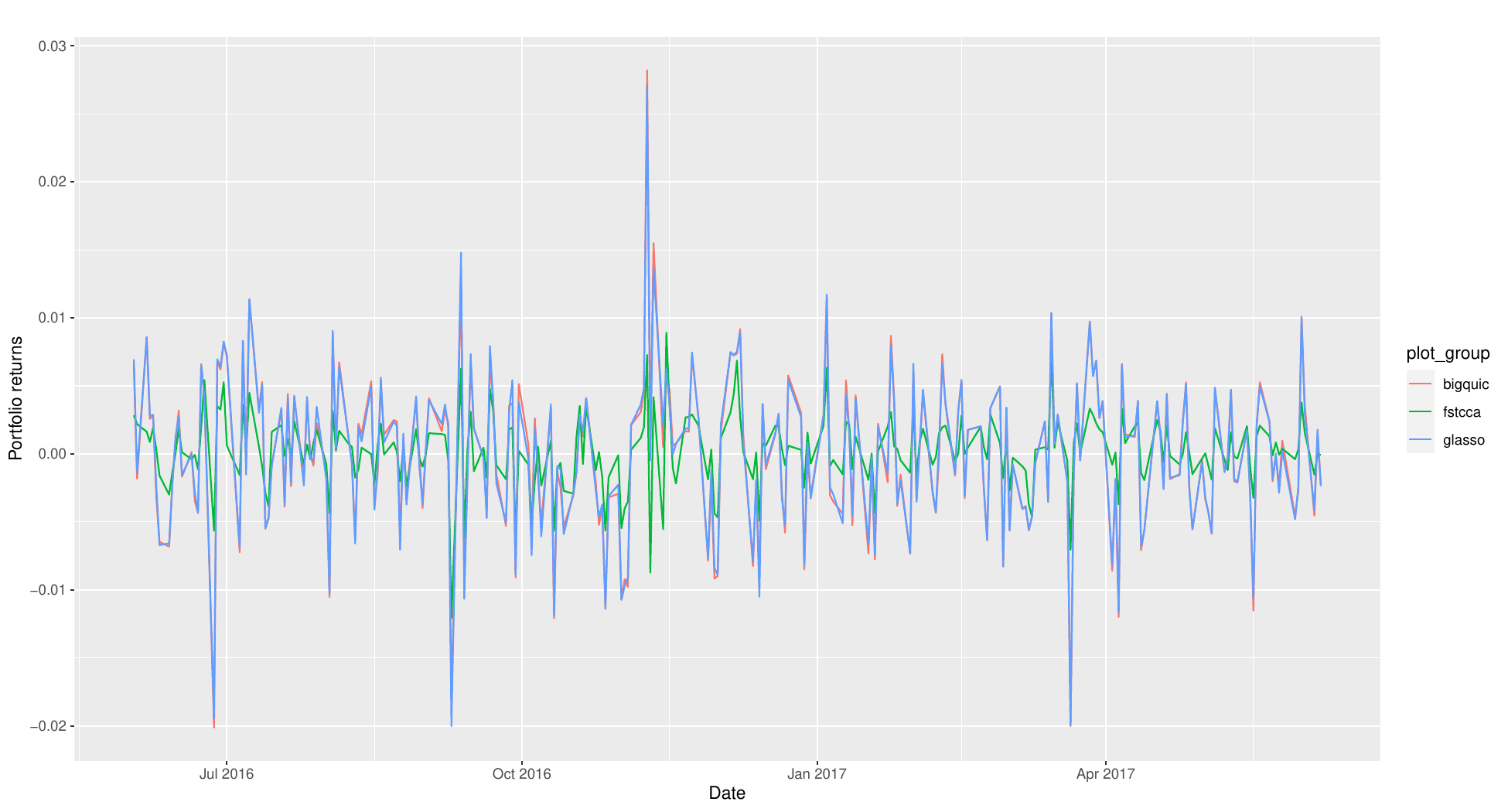}
	\end{center}
	\caption{Daily returns corresponding to portfolios associated with various covariance estimation strategies for stock/ETF data from \cite{HugeStock}.}
 \label{fig:large_stock_data:return}
\end{figure}

%
%
%
%
\Urlmuskip=0mu plus 1mu\relax

\newpage

\renewcommand{\theequation}{S.\arabic{equation}}
\renewcommand{\thethm}{S.\arabic{thm}}
\renewcommand{\thelem}{S.\arabic{lem}}

\setcounter{equation}{0}
\setcounter{thm}{0}
\setcounter{lem}{0}

\bigskip
\begin{center}
{\large\bf SUPPLEMENTAL MATERIALS}
\end{center}

\section*{Section A: Proof of Theorem 1}

\noindent
{(a) We first obtain high probability bounds for $\left\| 
\hat{L}^D - \bar{L}\right\|_2$. While bounds for $\left\| 
\hat{\Omega}^D - \bar{\Omega}\right\|_2$ can be obtained by minor changes 
to the proof of \cite[Theorem 1]{RBLZ:2008}, substantial conceptual and 
algebraic modifications are needed to derive bounds for $\left\| 
\hat{L}^D - \bar{L}\right\|_2$, as presented below. 

Let $\bar{D} = \bar{D}_n$ denote the diagonal matrix whose diagonal 
entries match those of $\bar{\Sigma}_n = \bar{\Omega}_n^{-1}$. Let 
$\bar{R} = \bar{R}_n = \bar{D}^{-1/2} \bar{\Sigma}_n \bar{D}^{-1/2}$ 
denote the population correlation matrix, and let $\bar{U} = 
\bar{D}^{1/2} \bar{L}$ denote the Cholesky factor of $\bar{R}^{-1}$. At 
the sample level, let $\hat{D}$ denote the diagonal matrix whose diagonal 
entries match those of $S$, and $\hat{R} = \hat{D}^{-1/2} S 
\hat{D}^{-1/2}$. Consider the estimator $\hat{U}^D$ defined by 
\begin{equation} \label{corrdefn}
\hat{U}^D = \mbox{argmin}_{U \in \mathbb{L}_{G^D_\sigma}} \left\{ tr 
\left( UU^T \hat{R} \right) - 2 \log |U| \right\}. 
\end{equation}

\noindent
Using results in \cite{Roverato:2002}, it follows that 
\begin{equation} \label{cholfactor:corr}
\hat{U}^{D, >}_{\cdot j} = -\frac{(\hat{R}^{> j})^{-1} \hat{R}^{>}_{\cdot j}}{\sqrt{\hat{R}_{jj} - (\hat{R}^{>}_{\cdot j})^T (\hat{R}^{> j})^{-1} 
		\hat{R}^{>}_{\cdot j}}} \mbox{ and } \hat{U}^D_{jj} = \frac{1}{\sqrt{\hat{R}_{jj} - (\hat{R}^{>}_{\cdot j})^T (\hat{R}^{> j})^{-1} \hat{R}^{>}_{\cdot j}}}.  
\end{equation}

\noindent
It follows from Equation (1) of the main paper and (\ref{cholfactor:corr}) that 
$\widehat{L}^D = \hat{D}^{-1/2} \hat{U}^D$. Using a strategy in 
\cite{RBLZ:2008}, we will first analyze $\left\| \hat{U}^D - \bar{U} 
\right\|$, and leverage this analysis to bound $\left\| \hat{L}^D - 
\bar{L} \right\|$. 

Let $\mathcal{L}^*_{G_\sigma^D}$ denote the space of lower triangular 
matrices whose $(i,j)^{th}$ entry is $0$ whenever $(i,j) \notin 
E_\sigma^D$ (diagonal entries not restricted to be positive). Consider 
the function 
\begin{eqnarray}
G(\Delta) 
&=& tr((\bar{U}+\Delta)(\bar{U}+\Delta)^T \hat{R}) - 2 
\log|\bar{U}+\Delta| - \left( tr((\bar{U}\bar{U}^T \hat{R}) - 2 
\log|\bar{U}| \right) \nonumber\\
&=& tr \left( \Delta^T \hat{R} \Delta \right) + 2 tr \left( \Delta^T 
\hat{R} \bar{U} \right) - 2 \sum_{i=1}^p \log \left( 1 + 
\frac{\Delta_{ii}}{\bar{U}_{ii}} \right). \label{gform}
\end{eqnarray}

\noindent
It is implicitly assumed that $\Delta_{ii} > -\bar{U}_{ii}$ for every $1 
\leq i \leq p$ so that $G(\Delta)$ is well-defined. Note that $G$ is a 
convex function, and is minimized at $\hat{\Delta} = \hat{U}^D - 
\bar{U}$. Hence, $G(\hat{\Delta}) \leq 0$. Let $\epsilon_n = \sqrt{(|E^D_\sigma|+1) \log p/n}$. We will now show that there exists a constant $M > 0$ such that 
$$
\inf \left\{ \Delta \in \mathcal{L}^*_{G_\sigma^D}: \; \|\Delta\|_F = 
M \epsilon_n \right\} > 0
$$

\noindent
It would then follow that the minimizer $\hat{\Delta}$ satisfies 
$\|\hat{\Delta}\|_2 \leq \|\hat{\Delta}\|_F \leq M \epsilon_n$. With this 
goal in mind, fix arbitrarily $\Delta \in \mathcal{L}^*_{G_\sigma^D}$ 
such that $\|\Delta\|_F = M \epsilon_n$. Using (\ref{gform}) and the 
inequality $\log(1+x) \leq x$ for $x > -1$, we get 
\begin{eqnarray}
G(\Delta) 
&\geq& tr \left( \Delta^T \hat{R} \Delta \right) + 2 tr \left( \Delta^T 
\hat{R} \bar{U} \right) - 2 \sum_{i=1}^p \frac{\Delta_{ii}}{\bar{U}_{ii}} \nonumber\\
&=& tr \left( \Delta^T \hat{R} \Delta \right) + 2 tr \left( \Delta^T 
(\hat{R}-\bar{R}) \bar{U} \right) + 2 tr \left( \Delta^T 
\bar{R} \bar{U} \right) - 2 \sum_{i=1}^p \frac{\Delta_{ii}}{\bar{U}_{ii}} \nonumber\\
&=& tr \left( \Delta^T \hat{R} \Delta \right) + 2 tr \left( 
(\hat{R}-\bar{R}) \bar{U} \Delta^T \right) + 2 tr \left( \Delta^T 
(\bar{U}^{-1})^T \right) - 2 \sum_{i=1}^p \frac{\Delta_{ii}}{\bar{U}_{ii}} \label{lb1}
\end{eqnarray}

\noindent
Since $\Delta \in \mathcal{L}^*_{G_\sigma^D}$, and $\bar{U}$ is lower 
triangular, it follows that 
\begin{equation} \label{lb2}
2 tr \left( \Delta^T 
(\bar{U}^{-1})^T \right) - 2 \sum_{i=1}^p \frac{\Delta_{ii}}{\bar{U}_{ii}}
= 0. 
\end{equation}

\noindent
Also, since $\sigma$ is a perfect vertex elimination scheme for the 
decomposable graph $G_\sigma^D$ and $\bar{U} \in 
\mathcal{L}_{G_\sigma^D}$, it follows that $(\bar{U} \Delta^T)_{ij} = 0$ 
if $(i,j) \notin E^D_\sigma$. Also, by Assumptions A1 and A2, and 
\cite[Lemma 3]{bckllvnrec}, there exists a constant 
$\tilde{K}$ such that 
\begin{equation} \label{lb3}
\max_{1 \leq i,j \leq p} |\hat{R}_{ij} - \bar{R}_{ij}| \leq \tilde{K} \sqrt{\frac{\log p}{n}} \mbox{ and } \max_{1 \leq i,j \leq p} |S_{ij} - \bar{\Sigma}_{ij}| \leq \tilde{K} \sqrt{\frac{\log p}{n}}
\end{equation}

\noindent
on an event $C_n$ where $P(C_n) \rightarrow 1$ as $n \rightarrow \infty$. 
Since the diagonal entries of $\hat{R} - \bar{R}$ are zero, it follows 
that on $C_n$ 
\begin{eqnarray}
|tr \left( (\hat{R}-\bar{R}) \bar{U} \Delta^T \right)| 
&=& \left| \sum_{(i,j) \in E_\sigma^D} (\hat{R}-\bar{R})_{ij} 
(\bar{U} \Delta^T)_{ij} \right| \nonumber\\
&\leq& \tilde{K} \sqrt{\frac{\log p}{n}} \sqrt{|E_\sigma^D|} \|\Delta 
\bar{U}^T\|_F \nonumber\\
&\leq& \frac{\tilde{K}}{\delta} \sqrt{\frac{|E_\sigma^D| \log p}{n}} 
\|\Delta\|_F \nonumber\\
&\leq& \frac{\tilde{K}}{M \delta} \|\Delta\|_F^2. \label{lb4}
\end{eqnarray}

\noindent
Finally, restricting to $C_n$, using $(\Delta \Delta^T)_{ij} = 0$ for 
$(i,j) \notin E^D_\sigma$ and Assumption A1, we get 
\begin{eqnarray}
tr \left( \Delta^T \hat{R} \Delta \right) 
&\geq& tr \left( \Delta^T \bar{R} \Delta \right) - |tr \left( \Delta^T 
(\hat{R} - \bar{R}) \Delta \right)| \nonumber\\
&=& tr \left( \Delta^T \bar{R} \Delta \right) - |tr \left( \Delta^T 
(\hat{R} - \bar{R}) \Delta \Delta^T \right)| \nonumber\\
&\geq& \frac{1}{\delta} \|\Delta\|_F^2 - \left| \sum_{(i,j) \in 
E_\sigma^D} (\hat{R}-\bar{R})_{ij} (\Delta \Delta^T)_{ij} \right| 
\nonumber\\
&\geq& \frac{1}{\delta} \|\Delta\|_F^2 - \tilde{K} \sqrt{\frac{\log 
p}{n}} \sqrt{|E_\sigma^D|} \|\Delta \|_F^2 \nonumber\\
&\geq& \frac{1}{2 \delta} \|\Delta \|_F^2 \label{lb5}
\end{eqnarray}

\noindent
for large enough $n$. It follows by (\ref{lb1}), (\ref{lb2}), (\ref{lb4}), (\ref{lb5}) that $G(\Delta) > \frac{1}{4 \delta} \|\Delta \|_F^2$ for $M = 8\tilde{K}/\delta$. Hence 
\begin{equation} \label{lb6}
P \left( \left\| \hat{U}^D - \bar{U} \right\|_2 \leq \frac{8 \tilde{K}}{\delta} \epsilon_n \right) \geq P(C_n) 
\end{equation}

\noindent
for large enough $n$. If we restrict to $C_n$, it follows from Assumption 
A1, (\ref{lb3}) and (\ref{lb6}) that 
\begin{eqnarray}
\left\| \hat{L}^D - \bar{L} \right\|_2 
&=& \left\| \hat{D}^{-1/2} \hat{U}^D - \bar{D}^{-1/2} \bar{U} \right\|_2 
\nonumber\\
&\leq& \|\hat{D}\|_2 \left\| \hat{U}^D - \bar{U} \right\|_2 + \left\| 
\hat{D}^{-1/2} - \bar{D}^{1/2} \right\|_2 \|\bar{U}\|_2 \nonumber\\
&\leq& \frac{16 \tilde{K}}{\delta^2} \epsilon_n + 
\frac{2 \tilde{K}}{\delta^3} \sqrt{\frac{\log p}{n}} \leq \frac{18 \tilde{K}}{\delta^3} \epsilon_n \label{lb7}
\end{eqnarray}

\noindent
for large enough $n$. The above bound was obtained in terms of $|E_\sigma^D|$. Using the closed from representation of $\hat{L}^D$ in 
Equation (1) of the main paper, we will now derive another bound for $\|\hat{L}^D - 
\bar{L} \|_2$ in terms of $a^D = a^D_n$, where $a^D_n$ is the product of 
$\max_{1 \leq j \leq p_n} n_j+1$ and $\max_{1 \leq i \leq p_n} r_i+1$. Here 
$n_j = \{i: 1 \leq j < i \leq p, (i,j) \in E_\sigma^D\}$ and $r_i = \{j: 
1 \leq j < i \leq p, (i,j) \in E_\sigma^D\}$. For this purpose, we will 
first bound the $\ell_2$-norm of the $j^{th}$ column of $\hat{L}^D - 
\bar{L}$ for an arbitrary $1 \leq j \leq p$. Using Equation (1) of the main paper, 
Assumption A1, $\bar{\Omega} \in \mathbb{P}_{G_\sigma^D}$, and the 
triangle inequality, we get 
\begin{eqnarray}
\|\hat{L}^D_{.j} - \bar{L}_j\|_2 
&\leq& \hat{L}_{jj} \left\| (S^{>j})^{-1} S^>_{.j} - 
(\bar{\Sigma}^{>j})^{-1} \bar{\Sigma}^>_{.j} \right||_2 + 
\left\| (\bar{\Sigma}^{>j})^{-1} \bar{\Sigma}^>_{.j} \right\|_2 
|\hat{L}_{jj} - \bar{L}_{jj}|. \label{lb8}
\end{eqnarray}

\noindent
To bound the above expression, it is key to bound $\|(S^{\geq j})^{-1} - 
(\bar{\Sigma}^{\geq j})^{-1}\|_2$. Here $A^{\geq j} = ((A_{kl}))_{k,l 
\geq j, (k,j),(l,j) \in E^D_\sigma}$ for any matrix $A$. For any ${\bf u} 
\in \mathbb{R}^{n_j}$ with $\|{\bf u}\|_2 = 1$, let $U$ be an $(n_j+1) 
\times (n_j+1)$ orthogonal matrix with first row ${\bf u}^T$. Then 
$US^{\geq j}U^T$ and $U\bar{\Sigma}^{\geq j}U^T$ are respectively the 
sample and population covariance matrix of $U{\bf Y}^{\geq j}_1, 
U{\bf Y}^{\geq j}_2, \cdots, U{\bf Y}^{\geq j}_n$. Note that the 
eigenvalues of $U\bar{\Sigma}^{\geq j}U^T$ are still uniformly bounded 
above and below by $\delta^{-1}$ and $\delta$ respectively, and 
${\bf u}^T S^{\geq j} {\bf u}$ is the first diagonal entry of 
$US^{\geq j}U^T$ (same with ${\bf u}^T \bar{\Sigma}^{\geq j} {\bf u}$. It 
follows by \cite[Lemma 3]{bckllvnrec} that 
$$
P \left( \left| {\bf u}^T S^{\geq j} {\bf u} - {\bf u}^T 
\bar{\Sigma}^{\geq j} {\bf u} \right| \leq \nu \right) \geq 1 - c_1 \exp(-c_2 n \nu^2) \mbox{ whenever } |\nu| < \kappa, 
$$

\noindent
where $c_1, c_2, \kappa$ only depend on $\delta$. Using the covering 
arguments in \cite[Lemma 5.2]{Vershynin:2012}, it follows that for any 
constant $K_1$
\begin{eqnarray*}
& & P \left( \sup_{\|{\bf u}\|_2 = 1} \left| {\bf u}^T S^{\geq j} {\bf u} - {\bf u}^T 
\bar{\Sigma}^{\geq j} {\bf u} \right| \leq K_1 \sqrt{\frac{(n_j+1)\log p}{n}} \right)\\
&\geq& 1 - \exp( - K_1^2 (n_j+1) \log p + 2(n_j+1) \log 21)
\end{eqnarray*}

\noindent
for large enough $n$. Since the above holds for every $1 \leq j \leq p$, 
the union-sum inequality can be used to establish the existence of an 
event $\tilde{C}_n$ such that $P(\tilde{C}_n) \rightarrow 1$ as $n 
\rightarrow \infty$, and on the event $\tilde{C}_n$ 
\begin{equation} \label{lb9}
\left\| S^{\geq j} - \bar{\Sigma}^{\geq j} \right\|_2 
\leq K_1 \sqrt{\frac{(n_j+1)\log p}{n}} \; \forall \; 1 \leq j \leq p 
\end{equation}

\noindent
for an appropriate constant $K_1$. Since $\max_{1 \leq j \leq p} n_j + 1 
= o(n/\log p)$, it follows from Assumption A1 and (\ref{lb9}) that on the 
event $\tilde{C}_n$ 
\begin{equation} \label{lb10}
\left\| (S^{\geq j})^{-1} - (\bar{\Sigma}^{\geq j})^{-1} \right\|_2 
\leq K_2 \sqrt{\frac{(n_j+1)\log p}{n}} \; \forall \; 1 \leq j \leq p
\end{equation}

\noindent
for an appropriate constant $K_2$. Using (\ref{lb10}) along with the fact 
that the first diagonal entries of $(S^{\geq j})^{-1}$ and 
$(\bar{\Sigma}^{\geq j})^{-1}$ are $\hat{L}_{jj}^2$ and $\bar{L}_{jj}^2$ 
respectively, we get 
$$
\left| \hat{L}_{jj}^2 - \bar{L}_{jj}^2 \right| \leq K_2 
\sqrt{\frac{(n_j+1)\log p}{n}}
$$

\noindent
for $1 \leq j \leq p$ on $\tilde{C}_n$. Using Assumption A1 along with 
$|\hat{L}_{jj} - \bar{L}_{jj}| = |\hat{L}_{jj}^2 - \bar{L}_{jj}^2|/|\hat{L}_{jj} + \bar{L}_{jj}|$, we get 
\begin{equation} \label{lb11}
\left| \hat{L}_{jj} - \bar{L}_{jj} \right| \leq \frac{K_2}{\delta} 
\sqrt{\frac{(n_j+1)\log p}{n}}
\end{equation}

\noindent
Since $\hat{L}_{ii}^2 (S^{>j})^{-1} S^>_{.j}$ is the first column of 
$(S^{\geq j})^{-1}$ (sans the diagonal entry), and $\bar{L}_{ii}^2 
(\bar{\Sigma}^{>j})^{-1} \bar{\Sigma}^>_{.j}$ is the first column of 
$(\bar{\Sigma}^{\geq j})^{-1}$, it follows from (\ref{lb8}), (\ref{lb10}) 
and (\ref{lb11}) that 
\begin{equation} \label{lb12}
\|\hat{L}^D_{.j} - \bar{L}_j\|_2 \leq K_3 \sqrt{\frac{(n_j+1)\log p}{n}} 
\end{equation}

\noindent
for every $1 \leq j \leq p$ on $\tilde{C}_n$. For any ${\bf v} \in 
\mathbb{R}^p$ with $\|{\bf v}\|_2 = 1$, we get by (\ref{lb12}) and the 
Cauchy-Schwarz inequality that 
\begin{eqnarray*}
\|(\hat{L}^D - \bar{L}){\bf v}\|_2^2 
&=& \sum_{i=1}^p \left( \sum_{j \leq i: i=j \; \mbox{or} \; (i,j) \in 
E_\sigma^D} (\hat{L}^D - \bar{L})_{ij} v_j \right)^2\\
&\leq& \sum_{i=1}^p (r_i+1) \sum_{j \leq i: i=j \; \mbox{or} \; (i,j) \in 
E_\sigma^D} (\hat{L}^D - \bar{L})_{ij}^2 v_j^2\\
&\leq& \left( \max_{1 \leq i \leq p} r_i + 1 \right) \sum_{j=1}^p v_j^2 
\left( \sum_{i \geq j: i=j \; \mbox{or} \; (i,j) \in E_\sigma^D} 
(\hat{L}^D - \bar{L})_{ij}^2 \right)\\
&=& \left( \max_{1 \leq i \leq p} r_i + 1 \right) \sum_{j=1}^p v_j^2 
\|\hat{L}^D_{.j} - \bar{L}_j\|_2^2\\
&\leq& K_3^2 \frac{a^D_n \log p}{n} \|{\bf v}\|_2^2
\end{eqnarray*}

\noindent
on $\tilde{C}_n$. It follows by the above inequality and (\ref{lb7}) that 
\begin{equation} \label{lb13}
\|\hat{L}^D - \bar{L}\|_2 \leq \max \left( K_3, \frac{18 \tilde{K}}{\delta^3} \right) \sqrt{\frac{\min(a^D_n, |E_\sigma^D|+1) \log p}{n}}
\end{equation}

\noindent
on $C_n \cap \tilde{C}_n$. Note that $P(C_n \cap \tilde{C}_n) \rightarrow 
1$ as $n\rightarrow \infty$. This establishes part (a). 

\medskip

\noindent
(b) Let $c = |E_\sigma^D \setminus E_\sigma|$. In other words, $c$ is the number of fill-in entries. Consider row-wise traversal of a lower triangular matrix as described in Step II of the CCA algorithm: start with the second row and move from entries with the lowest column index to the highest column index before the diagonal. We order the fill-in entries based on when they are encountered in this traversal. We define functions $\mathcal{U}$ and $\mathcal{V}$ from $\{1,2, \cdots, c\}$ to $\{1,2, \cdots, p\}$ such that $(\mathcal{U}(r), \mathcal{V}(r))$ is the $r^{th}$ fill-in entry, for $1 \leq r \leq c$. For $0 \leq r \leq c$, let $\hat{L}^{(r)}$ denote the intermediate matrix which is obtained by updating the first $r$ fill in entries as in in Algorithm 1 of the main paper, and $\hat{\Omega}^{(r)} = \hat{L}^{(r)} (\hat{L}^{(r)})^T$. In particular, $\hat{L}^{(0)} = \hat{L}^D$ while 
$\hat{L}^{(c)} = \widehat{L}$ (the {\it CCA} estimator). 

Henceforth, we will restrict to the event $C_n \cap \tilde{C}_n$. Let 
$$
\tilde{\epsilon}_n = \max(K_3, 18 \tilde{K} \delta^{-3})
\sqrt{\frac{\min(a^D_n, |E_\sigma^D|+1) \log p}{n}},
$$

\noindent
and let $N \in \mathbb{N}$ be such that $\tilde{\epsilon}_n 
\left( 1+\max_{1 \leq r \leq c} g^D (r) \right) < \frac{\sqrt{\delta}}{2}$ for $n > N$. Note 
that the matrices $\hat{L}^{(1)}$ and $\hat{L}^{(0)}$ only differ 
in the $(\mathcal{U}(1),\mathcal{V}(1))^{th}$ entry. Hence 
\begin{align*}
\norm{\hat{L^{(1)}} - \bar{L}}_2 &\leq \norm{\hat{L^{(1)}} - \hat{L^{(0)}}}_2 + \norm{\hat{L^{(0)}} - \bar{L}}_2 \\
&= \abs{\hat{L}^{(1)}_{\mathcal{U}(1),\mathcal{V}(1)} - \hat{L}^{(0)}_{\mathcal{U}(1),\mathcal{V}(1)}} + \norm{\hat{L^{(0)}} - \bar{L}}_2 \\
&= \abs{\hat{L}^{(0)}_{\mathcal{U}(1),\mathcal{V}(1)} + \frac{\sum _{k < \mathcal{V}(1)} \hat{L}^{(0)}_{\mathcal{U}(1),k} \hat{L}^{(0)}_{\mathcal{V}(1),k}}{\hat{L}^{(0)}_{\mathcal{V}(1),\mathcal{V}(1)}}} + \norm{\hat{L}^{(0)} - \bar{L}}_2 \\
&\leq  \frac{ \abs{  \hat{\Omega}^{(0)}_{\mathcal{U}(1),\mathcal{V}(1)} } }{ \abs{ \hat{L}^{(0)}_{\mathcal{V}(1),\mathcal{V}(1)} } } + \norm{\hat{L}^{(0)} - \bar{L}}_2 \\
&=  \frac{ \abs{  \hat{\Omega}^{(0)}_{\mathcal{U}(1),\mathcal{V}(1)}  - \bar{\Omega}_{\mathcal{U}(1),\mathcal{V}(1)} }}{ \abs{ \hat{L}^{(0)}_{\mathcal{V}(1),\mathcal{V}(1)} } } + \norm{\hat{L}^{(0)} - \bar{L}}_2
\end{align*}
Note that $\norm{\bar{\Omega}}_2 = \norm{ \bar{L} \bar{L}^T }_2 = \norm{\bar{L}}_2^2 \leq \frac{1}{\delta}$. It follows that
\begin{align*}
\norm{ \hat{ \Omega }^{ (0) } - \bar{ \Omega } }_2 &= \norm{  \hat{ L }^{ (0) }  (\hat{ L }^{ (0) })^T - \bar{ L } \bar{ L } ^T}_2 \\ 
&\leq \norm{  \hat{ L }^{ (0) }}_2  \norm{\hat{ L }^{ (0) }  - \bar{ L }}_2  +  \norm{\bar{ L }}_2 \norm{ \hat{ L }^{ (0) }  - \bar{ L } }_2 \\ 
&= (\norm{  \hat{ L }^{ (0) }}_2  +  \norm{\bar{ L }}_2) \norm{ \hat{ L }^{ (0) }  - \bar{ L } }_2 \\ 
&\leq (\norm{\hat{ L }^{ (0) } - \bar{L}} +  2 \norm{\bar{ L }}_2) \norm{ \hat{ L }^{ (0) }  - \bar{ L } }_2 \\ 
&\leq \frac{3}{\sqrt{\delta}} \norm{ \hat{ L }^{ (0) }  - \bar{ L } }_2
\end{align*}

\noindent
on $C_n \cap \tilde{C}_n$ for large enough $n$. The last inequality 
follows from (\ref{lb13}) and $\sqrt{\frac{\min(a^D_n, |E_\sigma^D|+1) 
\log p}{n}} = o(1)$. Also by (\ref{lb13}), as ${\bar{L}_{ii}} \geq 
{\sqrt{\delta}}$ and $\abs{ \hat{L}^{(0)}_{ii}} \geq \abs{ 
\bar{L}_{ii}} - \abs{ \bar{L}_{ii} - \hat{L}^{(0)}_{ii}}$, 
we have $\abs{ \hat{L}^{(0)}_{ii}} \geq \frac{\sqrt{\delta}}{2}$ for 
every $1 \leq i \leq p$ on $C_n \cap \tilde{C}_n$ for large enough $n$. 
It follows that,
\begin{equation} \label{lb14}
\abs{\hat{L}^{(1)}_{\mathcal{U}(1),\mathcal{V}(1)} - \hat{L}^{(0)}_{\mathcal{U}(1),\mathcal{V}(1)}} \leq \frac{6}{\delta} \tilde{\epsilon}_n = g^D (1) 
\end{equation}

\noindent
for $n > N$. We will now proceed by induction. Suppose 
$$
\abs{\hat{L}^{(\tilde{r})}_{\mathcal{U}(\tilde{r}),\mathcal{V}(\tilde{r})} - \hat{L}^{(0)}_{\mathcal{U}(\tilde{r}),\mathcal{V}(\tilde{r})}} \leq g^D(\tilde{r}) \tilde{\epsilon}_n
$$

\noindent
for $1 \leq \tilde{r} \leq r-1$. Note that 
\begin{eqnarray*}
\abs{\hat{L}^{({r})}_{\mathcal{U}({r}),\mathcal{V}({r})} - \hat{L}^{(0)}_{\mathcal{U}({r}),\mathcal{V}({r})}} 
&=& \abs{\hat{L}^{(0)}_{\mathcal{U}(r),\mathcal{V}(r)} + \frac{\sum _{k < \mathcal{V}(r)} \hat{L}^{(r-1)}_{\mathcal{U}(r),k} \hat{L}^{(r-1)}_{\mathcal{V}(r),k}}{\hat{L}^{(0)}_{\mathcal{V}(r),\mathcal{V}(r)}}}. 
\end{eqnarray*}

\noindent
Note that the entries of $\hat{L}^{(r-1)}$ and $\hat{L}^{(0)}$ match 
except for the first $r-1$ fill-in entries. Recall that $\tilde{r} \in 
\mathcal{F}^D_r$ if and only if $(\mathcal{U}(\tilde{r}),\mathcal{V}
(\tilde{r})) = (\mathcal{U}(r),k)$ and $((\mathcal{V}(r),k) \in 
E^D_\sigma$ for some $k < \mathcal{V}(r)$ or $(\mathcal{U}(\tilde{r}),
\mathcal{V}(\tilde{r})) = (\mathcal{V}(r),k)$ and $(\mathcal{U}(r),k) \in 
E^D_\sigma$ for some $k < \mathcal{V}(r)$. By the induction hypothesis, 
(\ref{lb13}), and the choice of $N$, it follows that 
$|\hat{L}^{(\tilde{r})}_{ij}| \leq \frac{3}{2 \sqrt{\delta}}$ whenever $i 
> j$, and $|\hat{L}^{(\tilde{r})}_{ij}| \geq \frac{\sqrt{\delta}}{2}$ 
whenever $i=j$. Using $|ab-a_0b_0| \leq |a||b-b_0| + |b_0||a-a_0|$ (in 
case there is a $k < \mathcal{V}(r)$ such that $(\mathcal{U}(r),k)$ and 
$(\mathcal{V}(r),k)$ are both fill-in entries) along with the above 
observations we get 
\begin{eqnarray*}
& & \abs{\hat{L}^{({r})}_{\mathcal{U}({r}),\mathcal{V}({r})} - 
\hat{L}^{(0)}_{\mathcal{U}({r}),\mathcal{V}({r})}}\\
&\leq& \abs{\hat{L}^{(0)}_{\mathcal{U}(r),\mathcal{V}(r)} + \frac{\sum _{k < \mathcal{V}(r)} \hat{L}^{(0)}_{\mathcal{U}(r),k} \hat{L}^{(0)}_{\mathcal{V}(r),k}}{\hat{L}^{(0)}_{\mathcal{V}(r),\mathcal{V}(r)}}} + \frac{3}{\delta} \sum_{\tilde{r} \in \mathcal{F}^D_r} \abs{\hat{L}^{(\tilde{r})}_{\mathcal{U}(\tilde{r}),\mathcal{V}(\tilde{r})} - \hat{L}^{(0)}_{\mathcal{U}(\tilde{r}),\mathcal{V}(\tilde{r})}}\\
&\leq& \frac{ \abs{  \hat{\Omega}^{(0)}_{\mathcal{U}(r),\mathcal{V}(r)}  - \bar{\Omega}_{\mathcal{U}(r),\mathcal{V}(r)} }}{ \abs{ \hat{L}^{(0)}_{\mathcal{V}(r),\mathcal{V}(r)} } } + \frac{3}{\delta} \sum_{\tilde{r} \in \mathcal{F}^D_r} g^D (\tilde{r}) \tilde{\epsilon}_n\\
&\leq& \frac{3}{\delta} \left( 2 + \sum_{\tilde{r} \in \mathcal{F}^D_r} g^D (\tilde{r}) \right) \tilde{\epsilon}_n = g^D (r) \tilde{\epsilon}_n. 
\end{eqnarray*}

\noindent
By induction, it follows that 
$$
\abs{\hat{L}^{({r})}_{\mathcal{U}({r}),\mathcal{V}({r})} - 
\hat{L}^{(0)}_{\mathcal{U}({r}),\mathcal{V}({r})}} \leq g^D (r) 
\tilde{\epsilon}_n \; \; \forall \; \; 1 \leq r \leq c. 
$$

\noindent
This implies 
$$
\norm{\hat{L} - \bar{L}^D}_{max} = \norm{\hat{L}^{(c)} - \hat{L}^{(0)}}_{max} \leq \tilde{\epsilon}_n \max_{1 \leq r \leq c} g^D (r), 
$$

\noindent
Using a similar argument as the one right before (\ref{lb13}), we get 
$$
\norm{\widehat{L} - \widehat{L}^D}_2 \leq \tilde{\epsilon}_n \sqrt{\tilde{a}^D_n} 
\max_{1 \leq r \leq c} g^D (r) 
$$

\noindent
where $\tilde{a}^D_n$ is the product of the maximum number of fill-in 
entries in any given column and the maximum number of fill-in entries in 
any given row. It follows that on $C_n \cap \tilde{C}_n$ 
\begin{equation} \label{lb15}
\norm{\hat{L} - \bar{L}}_2 \leq \tilde{\epsilon}_n \left( 
\sqrt{\tilde{a}^D_n} \max_{1 \leq r \leq c} g^D (r) + 1 \right) \leq K 
\sqrt{\frac{s_{CCA} \log p}{n}} 
\end{equation}

\noindent
for an appropriate constant $K$. Finally, note that 
\begin{eqnarray*}
	\norm{\widehat{\Omega} - \bar{\Omega}}_2 
	&=& \norm{\widehat{L} \widehat{L}^T - \bar{L} \bar{L}^T}_2\\
	&=& \norm{\widehat{L} \widehat{L}^T - \bar{L} \widehat{L}^T + \bar{L} \widehat{L}^T - \bar{L} \bar{L}^T}_2\\
	&\leq& \norm{\widehat{L} - \bar{L}}_2 \norm{\widehat{L}}_2 + \norm{\bar{L}}_2 \norm{\widehat{L} - \bar{L}}_2\\
	&\leq& \norm{\widehat{L} - \bar{L}}_2^2 + 2 \norm{\bar{L}}_2 \norm{\widehat{L} - \bar{L}}_2\\
    &\leq& K' \sqrt{\frac{s_{CCA} \log p}{n}}
\end{eqnarray*}

\noindent
on $C_n \cap \tilde{C}_n$ for an appropriate constant $K'$. The last 
inequality follows from (\ref{lb15}), Assumption A1 and Assumption A2.} 
\hfill$\Box$

\section*{Section B: Weak convergence with fixed $p$}

\noindent
We start by establishing two lemmas which imply that $\hat{\Omega}$ is a 
differentiable function of $\Omega$. 
\begin{lem}
	The function $\phi_{1, G_\sigma^D}: \mathbb{P}^+ \rightarrow \mathbb{P}_{G_\sigma^D}$ defined by 
	$$
	\phi_{1, G_\sigma^D} (A) = \sum_{C \in \mathbb{C}_\sigma^D} \left[ (A^{-1})_C \right]^0 - 
	\sum_{Sep \in \mathcal{S}_\sigma^D} \left[ (A^{-1})_{Sep} \right]^0 
	$$
	
	\noindent
	is differentiable. Also, if $A^{-1} \in \mathbb{P}_{G_\sigma^D}$, then $\phi_{1, G_\sigma^D} (A) = 
	A^{-1}$. 
\end{lem}

\noindent
{\it Proof:} The differentiability follows by repeated application of the differentiability 
of the matrix inverse for positive definite matrices. It is known that 
(see \cite{lrtzngphmd}) $\phi_{1, G_\sigma^D} (A)$ uniquely minimizes the function 
$$
f(K) = tr(KA) - |KA| 
$$

\noindent
for $K \in \mathbb{P}_{G_\sigma^D}$. Since $A^{-1}$ uniquely minimizes $f$ over the 
space of all positive definite matrices, if $A^{-1} \in \mathbb{P}_{G_\sigma^D}$, it 
follows that $\phi_{1, G_\sigma^D} (A) = A^{-1}$. 
\hfill$\Box$

\noindent
Let $\phi_{2, G_\sigma}: \mathbb{L}^+ \rightarrow \mathbb{L}_{G_\sigma}$ be defined as the function 
which takes $L \in \mathbb{L}^+$, and applies the last step of Algorithm 1 of the main paper (with the two {\it for} loops) to convert it to a matrix in $\mathcal{L}_{G_\sigma}$. 
\begin{lem}
	The function $\phi_{3, G_\sigma}: \mathbb{P}^+ \rightarrow \mathbb{P}_{G_\sigma}$ defined by 
	$$
	\phi_{3, G_\sigma^D} (A) = \phi_{2, G_\sigma} (L_A) \phi_{2, G_\sigma} (L_A)^T 
	$$
	
	\noindent
	is differentiable. Here $L_A$ denotes the Cholesky factor of $A$. Also, if $A \in 
	\mathbb{P}_{G_\sigma}$, then $\phi_{3, G_\sigma} (A) = A$. 
\end{lem}

\noindent
{\it Proof:} The transformation of a positive definite matrix $A$ to its Cholesky factor $L_A$ 
is clearly differentiable. It is clear from construction that for $i > j$, 
$(\phi_{2, G_\sigma} (L_A))_{ij}$ is a polynomial in $\left( \left( L_A \right)_{ij}, 
\left\{ \phi_{2, G_\sigma} (L_A)_{ik} \right\}_{k < j}, \left\{ \phi_{2, G_\sigma} (L_A)_{jk} 
\right\}_{k < j}, \frac{1}{\left( L_A \right)_{jj}} \right)$. Since all the diagonal entries of 
$\phi_{2, G_\sigma} (L_A)$ are the same as the diagonal entries of $L_A$, it follows that 
the entries of $\phi_{2, G_\sigma} (L_A)$ are polynomials in entries of $L_A$ and $\left\{ 
\frac{1}{\left( L_A \right)_{ii}} \right\}_{1 \leq i \leq p}$, which implies that all the entries of 
$\phi_{3, G_\sigma} (A)$ are polynomials in the entries of $L_A$ and $\left\{ 
\frac{1}{\left( L_A \right)_{ii}} \right\}_{1 \leq i \leq p}$. Hence, $\phi_{3, G_\sigma}$ is a 
differentiable transformation from $\mathbb{P}^+$ to $\mathbb{P}_{G_\sigma}$. 

Suppose $A \in \mathbb{P}_{G_\sigma}$. It follows that $L_A \in \mathcal{L}_{G_\sigma}$. 
We prove by induction that $\phi_{2, G_\sigma}$ does not alter $L_A$. This would clearly 
imply $\phi_{3, G_\sigma} (A) = A$. Note that all the diagonal entries of $L_A$ are not 
altered by $\phi_{2, G_\sigma}$. Suppose that all the entries of 
$\phi_{2, G_\sigma} (L_A)$ and $L_A$ are the same for the first $i-1$ rows, and the first 
$j-1$ entries of the $i^{th}$ row, for some $i > j$. If $(i,j) \in E_\sigma$, then 
by construction $\phi_{2, G_\sigma} (L_A)_{ij} = (L_A)_{ij}$. If $(i,j) \notin E_\sigma$, 
then by the induction hypothesis and the fact that $L_A \in \mathcal{L}_{G_\sigma}$, it 
follows that 
\begin{eqnarray*}
	\phi_{2, G_\sigma} (L_A)_{ij} = 
	&=& \frac{-\sum_{k=1}^{j-1} \phi_{2, G_\sigma} (L_A)_{ik} \phi_{2, G_\sigma} (L_A)_{jk}}
	{\phi_{2, G_\sigma} (L_A)_{jj}}\\
	&=& \frac{-\sum_{k=1}^{j-1} (L_A)_{ik} (L_A)_{jk}}{(L_A)_{jj}}\\
	&=& (L_A)_{ij}. 
\end{eqnarray*}

\noindent
The result follows by induction. \hfill$\Box$ 

\noindent
Using the above lemmas, we now establish weak convergence of the estimator corresponding to 
Algorithm 1 of the main paper.  
\begin{thm}
	Under the assumption in Equation 5 of the main paper, the estimator $\widehat{\Omega}$ obtained from 
	Algorithm 1 of the main paper is $\sqrt{n}$-consistent. In particular, $\sqrt{n}(\widehat{\Omega} - 
	\bar{\Omega})$ converges to a multivariate Gaussian limit as $n \rightarrow \infty$. 
\end{thm}

\noindent
{\it Proof:} Note that $\widehat{\Omega}^D = \phi_{1, G_\sigma} (S)$ and $\widehat{\Omega} = 
\phi_{3, G_\sigma} (\widehat{\Omega}^D)$. Hence, we get that $\widehat{\Omega} = 
\phi_{3, G_\sigma} \left( \phi_{1, G_\sigma} (S) \right)$. Since $\bar{\Omega} \in 
\mathbb{P}_{G_\sigma}$, it follows that $\bar{\Omega} = \phi_{3, G_\sigma} \left( 
\phi_{1, G_\sigma} (\bar{\Sigma}) \right)$. Using the multivariate Delta method along with 
these observations establishes the convergence of $\sqrt{n}(\widehat{\Omega} - 
\bar{\Omega})$ to a multivariate Gaussian limit as $n \rightarrow \infty$. \hfill$\Box$

\section*{Section C: Proof of Lemma 1}

\noindent
Note that the assumption that $n$ is greater than the largest clique size in $G_\sigma^D$ is required for the existence of 
$\widehat{\Omega}^D$, which in turn is needed for Algorithm 1 in the main paper. Let $\widehat{L}$ denote the 
Cholesky factor of $\widehat{\Omega}$. It is clear by construction that $\widehat{L} \in 
\mathbb{L}_{G_\sigma^D}$, and $\widehat{\Omega} \in \mathbb{P}_{G_\sigma^D}$. Note that if $i>j, (i,j) 
\notin E_\sigma^D \setminus E_\sigma$, then $\widehat{L}_{ij} = 
-\sum_{k=1}^{j-1} \widehat{L}_{ik} \widehat{L}_{jk}/\widehat{L}_{jj}$. Hence 
$$
\widehat{\Omega}_{ij} = (\widehat{L} \widehat{L}^T)_{ij} = \sum_{k=1}^j \widehat{L}_{ik} \widehat{L}_{jk} = 
\widehat{L}_{ij} \widehat{L}_{jj} + \sum_{k=1}^{j-1} \widehat{L}_{ik} \widehat{L}_{jk} = 0. 
$$

\noindent
It follows that $\widehat{L} \in \mathbb{L}_{G_\sigma}$ and $\widehat{\Omega} \in \mathbb{P}_{G_\sigma}$. 
\hfill$\Box$

\section*{Section D: Proof of Lemma 2}

\noindent
The fact that $\hat{L}$ is a minimizer of $h(\cdot)$ over the 
relevant set follows from the fact that $h$ is non-negative, $h(\hat{L}) 
= 0$, and Lemma 1 in the main paper. Suppose $\tilde{L}$ is another minimizer 
of $h$ over the relevant set. Then $h(\tilde{L}) = 0$, which implies that 
$\tilde{L}_{ij} = \widehat{L}^D_{ij} = \hat{L}_{ij}$ if $(i,j) \in 
E_\sigma$ or $i=j$. Suppose there are $c$ fill-in entries for $\hat{L}^D$ 
with row-column locations given by $(i_1, j_1), (i_2, j_2), \cdots, (i_c, 
j_c)$. The entries are ordered based on the traversal of $\widehat{L}^D$ 
described in Step II above. Since $(\tilde{L} \tilde{L}^T)_{i_1 j_1} = 
0$, it follows that 
$$
\tilde{L}_{i_1 j_1} = - \frac{\sum_{k=1}^{j_1-1} \tilde{L}_{i_1k} 
\tilde{L}_{j_1 k}}{\tilde{L}_{j_1 j_1}}. 
$$

\noindent
Since all entries on the RHS are not fill-in entries and 
$\widehat{\Omega}_{i_1 j_1} = (\widehat{L} \widehat{L}^T)_{i_1 j_1} = 0$, 
it follows that $\tilde{L}_{i_1 j_1} = \widehat{L}_{i_1 j_1}$. Similarly, 
if $c \geq 2$, note that 
$$
\tilde{L}_{i_r j_r} = - \frac{\sum_{k=1}^{j_r-1} \tilde{L}_{i_rk} 
\tilde{L}_{j_r k}}{\tilde{L}_{j_r j_r}}
$$

\noindent
for every $2 \leq r \leq c$. The entries on the RHS are either non fill-in entries or among the first $r-1$ fill-in entries. Hence, by a striaghtforward induction argument it follows that $\tilde{L}$ and 
$\widehat{L}$ share values even at the fill-in locations. \hfill$\Box$

\section*{Section E: Proof of Lemma 5}

\noindent
First let us assume $G$ is connected. Then 
\begin{eqnarray*}
2 p^2 |E_\sigma|  = p^2 \sum_{i=1}^p |\{j: (i,j) \in E_\sigma\}| \geq p^2 \times p = p^3. 
\end{eqnarray*}

\noindent
Similarly 
\begin{eqnarray*}
p^2 \sum_{U \in \mathcal{U}} |U| 
= p^2 \sum_{U \in \mathcal{U}} \sum_{i \in U} 1
= p^2 \sum_{i \in V} \sum_{U \in \mathcal{U}: \; i \in U} 1
\geq p^2 \sum_{i \in V} 1 = p^3. 
\end{eqnarray*}

\noindent
The result now follows by Table 1 in the main paper. For disconnected graphs, the inequalities above hold for each respective connected component. Since the overall complexity for all algorithms is obtained by 
adding the complexities over all connected components, the result follows for disconnected graphs as well. \hfill$\Box$

\section*{Section F: Proof of Lemma 6}

\noindent
Let $c_p = {{p}\choose{2}}$ and $d_p = p^{5/3}$ for convenience. Then 
$$
|\mathcal{A}_p| = 2^{c_p} - |\mathcal{A}_p^c| = 2^{c_p} - \sum_{k=0}^{d_p} {{c_p}\choose{k}} \geq 2^{c_p} - (d_p+1) 
{{c_p}\choose{d_p}} 
$$

\noindent
for $p \geq 100$, since ${{n}\choose{r}}$ increases in $r$ for $r < n/2$ and $d_p < c_p/2$ for $p \geq 100$. Using Stirling's 
approximation, it follows that 
$$
|\mathcal{A}_p| \geq 2^{c_p} - (d_p+1) \frac{c_p^{d_p}}{d_p!} \geq 2^{c_p} - C \sqrt{d_p} \left( \frac{e c_p}{d_p} \right)^{d_p} \geq 
2^{c_p} - C \sqrt{d_p} \left( e c_p \right)^{d_p}
$$

\noindent
where $C = \sqrt{2/\pi}$. Since $p^2/4 \leq c_p \leq p^2/2$ for $p \geq 100$, it follows that 
$$
\frac{|\mathcal{A}_p|}{2^{c_p}} \geq 1 - C \exp \left( \frac{5}{6} \log p + d_p + 2 d_p \log p - c_p \log 2 \right)  \geq 1 - C \exp \left( 3 d_p \log p - \frac{p^2 \log 2}{4} \right) \rightarrow 1 
$$

\noindent
as $p \rightarrow \infty$. This establishes part (a). For part (b), note that for any ordering $\sigma$ of the $p$ vertices, 
$\left| E_\sigma^D \right| \geq \left| E_\sigma \right| > p^{5/3}$. Hence, the {\it CCA} algorithm uses a direct computation of $\hat{L}^D$ 
(see equation (3) in the main paper) and has computational complexity $O(p^3)$. On the other hand, it can be seen using Jensen's 
inequality (similar to equation (3) in the main paper) that 
$$
\sum_{j=1}^p \tilde{n}_j^3 \geq p \left( \frac{1}{p} \sum_{j=1}^p \tilde{n}_j \right)^3 = \frac{|E_\sigma|^3}{p^2} \geq p^{3}. 
$$

\noindent
\hfill$\Box$

\section*{Section G: Proof of Lemma 7}

\noindent
Since $\hat{\Omega}_{32} = 0$, it follows that 
$$
\hat{L}_{32} = - \frac{\hat{L}_{21} \hat{L}_{31}}{\hat{L}_{22}}. 
$$

\noindent
Next, using $\hat{\Omega}_{43} = 0$ along with the above equality, it follows that 
$$
\hat{L}_{43} = - \frac{\hat{L}_{42} \hat{L}_{32}}{\hat{L}_{33}} = \frac{\hat{L}_{21} \hat{L}_{31} \hat{L}_{42}}{\hat{L}_{22} \hat{L}_{33}}. 
$$

\noindent
Sequentially extending these calculations to include all fill-in entries, we obtain 
\begin{equation} \label{CCA:fillin}
\hat{L}_{i,i-1} = (-1)^{i-2} \hat{L}_{i,i-2} \frac{\hat{L}_{21}}{\hat{L}_{22}} \prod_{j=3}^{i-1} \frac{\hat{L}_{j,j-2}}{\hat{L}_{jj}}
\end{equation}

\noindent
for $4 \leq i \leq p-1$. Since $\Omega_0 \in \mathbb{P}_{G_\sigma}$, identical calculations lead to 
\begin{equation} \label{truefillin}
\bar{L}_{32} = - \frac{\bar{L}_{21} \bar{L}_{31}}{\bar{L}_{22}} \mbox{ and } \bar{L}_{i,i-1} = (-1)^{i-2} \bar{L}_{i,i-2} \frac{\bar{L}_{21}}{\bar{L}_{22}} \prod_{j=3}^{i-1} \frac{\bar{L}_{j,j-2}}{\bar{L}_{jj}} 
\end{equation}

\noindent
for $4 \leq i \leq p-1$. In (\ref{CCA:fillin}) and (\ref{truefillin}), all fill-in entries of the respective Cholesky matrices are expressed in terms of the non fill-in entries of these matrices. Note that $a_n^D = O(1)$ in this setting. By repeating the arguments in the proof of part (a) of Theorem 1 after equation (S.10), there exists an appropriate positive constant $K^*$ such that 
$$
\norm{\widehat{L}^D - \bar{L}}_2 \leq \tilde{K}^* \sqrt{\frac{\log p}{n}} 
$$

\noindent
on an event $\tilde{D}_n$ with $\bar{P}(\tilde{D}_n) \rightarrow 1$ as $n 
\rightarrow \infty$. Let $\tilde{\delta} > 0$ be such that 
\begin{equation} \label{deltatilde}
\frac{2 \tilde{\delta}}{\delta} + \frac{4 \tilde{\delta}}{\delta^3} < \frac{1-\xi}{2}, 
\end{equation}

\noindent
where $\xi := \sup_{n \geq 1} \max_{3 \leq j \leq p-1} \frac{|\bar{L}_{j,j-2}|}{\bar{L}_{jj}} < 1$ by Assumption A3'. Since $\hat{L}^D_{i,j} = \hat{L}_{ij}$ for every $(i,j) \in E_\sigma$ (Step 2 of the CCA algorithm leaves the non fill-in entries invariant), it follows that 
$$
\max_{(i,j) \in E_\sigma} |\hat{L}_{ij} - \bar{L}_{ij}| < \tilde{\delta}
$$

\noindent
on an event $D_n$ such that $\bar{P}(D_n) \rightarrow 1$ as $n \rightarrow 
\infty$. Note by Assumption 1 and (\ref{deltatilde}) that on $D_n$
$$
\hat{L}_{jj} \geq \bar{L}_{jj} - |\hat{L}_{jj} - \bar{L}_{jj}| \geq \delta - \tilde{\delta} > \frac{\delta}{2} 
$$

\noindent
for $1 \leq j \leq p$, and 
$$
|\hat{L}_{ij}| \leq |\bar{L}_{ij}| + |\hat{L}_{ij} - \bar{L}_{ij}| \leq \frac{1}{\delta} + \tilde{\delta} < \frac{2}{\delta}
$$

\noindent
for every $(i,j) \in E_\sigma$. It follows that on $D_n \cap \tilde{D}_n$ 
\begin{eqnarray}
\left| \frac{\hat{L}_{j,j-2}}{\hat{L}_{jj}} - \frac{\bar{L}_{j,j-2}}{\bar{L}_{jj}} \right| 
&\leq& \left| \frac{\hat{L}_{j,j-2}}{\hat{L}_{jj}} - \frac{\hat{L}_{j,j-2}}{\bar{L}_{jj}} \right| + \left| \frac{\hat{L}_{j,j-2}}{\bar{L}_{jj}} - \frac{\bar{L}_{j,j-2}}{\bar{L}_{jj}} \right| \nonumber\\
&\leq& \frac{\hat{L}_{j,j-2}}{\hat{L}_{jj} \bar{L}_{jj}} \left| \hat{L}_{jj} - \bar{L}_{jj} \right| + \frac{|\hat{L}_{j,j-2} - \bar{L}_{j,j-2}|}{\bar{L}_{jj}} \nonumber\\
&\leq& \min \left( \frac{2 \tilde{\delta}}{\delta} + \frac{4 \tilde{\delta}}{\delta^3}, \left( \frac{2 \tilde{K}^*}{\delta} + \frac{4 \tilde{K}^*}{\delta^3} \right) \sqrt{\frac{\log p}{n}} \right) \nonumber\\
&<& \min \left( \frac{1-\xi}{2}, \left( \frac{2 \tilde{K}^*}{\delta} + \frac{4 \tilde{K}^*}{\delta^3} \right) \sqrt{\frac{\log p}{n}} \right), \label{diff:ratio}
\end{eqnarray}

\noindent
and 
\begin{equation} \label{boundratio}
\left| \frac{\hat{L}_{j,j-2}}{\hat{L}_{jj}} \right| \leq \left| \frac{\bar{L}_{j,j-2}}{\bar{L}_{jj}} \right| + \left| \frac{\hat{L}_{j,j-2}}{\hat{L}_{jj}} - \frac{\bar{L}_{j,j-2}}{\bar{L}_{jj}} \right| < \xi + \frac{1-\xi}{2} = \frac{1 + \xi}{2} 
\end{equation}

\noindent
for $3 \leq j \leq p$. By similar arguments as above, it can be shown that 
on $D_n \cap \tilde{D}_n$ 
\begin{equation} \label{boundextra}
\left| \hat{L}_{i,i-2} \frac{\hat{L}_{21}}{\hat{L}_{22}} - \bar{L}_{i,i-2} \frac{\bar{L}_{21}}{\bar{L}_{22}} \right| \leq \frac{8 \tilde{K}^*}{\delta^3} \sqrt{\frac{\log p}{n}} \mbox{ and } \left| \hat{L}_{i,i-2} \frac{\hat{L}_{21}}{\hat{L}_{22}} \right| \leq \frac{8}{\delta^3} 
\end{equation}

\noindent
for $3 \leq i \leq p$. Using the identity 
$$
\left| \prod_{j=1}^k t_j - \prod_{j=1}^k r_j \right| \leq |t_1 - r_1| \prod_{j=2}^k t_j + \sum_{i=2}^{k-1} \left( \prod_{j=1}^{i-1} t_j \right) \left( \prod_{j=i+1}^k r_j \right) |t_i - r_i| + \left( \prod_{j=1}^{k-1} t_j \right) |t_k - r_k|
$$

\noindent
along with (\ref{CCA:fillin}), (\ref{truefillin}), (\ref{diff:ratio}), 
(\ref{boundratio}), (\ref{boundextra}) and Assumption $A3^\prime$, we 
obtain 
$$
|\hat{L}_{i,i-1} - \bar{L}_{i,i-1}| \leq \left( \frac{8 \tilde{K}^*}{\delta^3} \left( \frac{1+\xi}{2} \right)^{i-3} + (i-3) \frac{8}{\delta^3} \left( \frac{2 \tilde{K}^*}{\delta} + \frac{4 \tilde{K}^*}{\delta^3} \right) \left( \frac{1+\xi}{2} \right)^{i-4} \right) \sqrt{\frac{\log p}{n}}
$$

\noindent
for every $3 \leq i \leq p-1$ on $D_n \cap \tilde{D}_n$. Since $(1+\xi)/2 < 1$, it follows that for an appropriate constant $\tilde{K}$ 
$$
\max_{(i,j) \in E^D_\sigma} |\hat{L}_{ij} - \bar{L}_{ij}| \leq \tilde{K} 
\sqrt{\frac{\log p}{n}} 
$$

\noindent
on $D_n \cap \tilde{D}_n$. Since $|E_\sigma| = p$ and $|E^D_\sigma| = 2p-3$, 
it follows that 
$$
\|\hat{L} - \bar{L}\|_F^2 \leq \tilde{K} \sqrt{\frac{(|E_\sigma|+1)\log p}{n}}
$$

\noindent
on $D_n \cap \tilde{D}_n$. The bound for $\|\hat{\Omega} - \bar{\Omega}\|_F$ now follows by similar arguments as those at the end of the proof of Theorem 1. \hfill$\Box$

\bibliographystyle{plainnat}
\bibliography{CCA_UNBLIND.bib}

\end{document}